\documentclass{article}
\usepackage{fullpage}
\DeclareMathAlphabet{\altmathcal}{OMS}{cmsy}{m}{n}
\usepackage{mathptmx}
\usepackage{color}
\usepackage{amsmath,amsbsy}
\usepackage{amsfonts}
\usepackage[square,numbers,sort&compress,super]{natbib}
\usepackage{url}
\usepackage{bbm}
\usepackage{mdwlist}
\usepackage{float}
\usepackage{graphicx}
\usepackage{courier}
\usepackage{fancyhdr}
\usepackage{hyperref}
\usepackage{booktabs}
\usepackage{pdflscape}
\usepackage[font=small,labelfont=bf]{caption}
\usepackage{amsmath}
\usepackage{textcomp}

\title{Statistical methods used to combine the effective reproduction number, $R(t)$, and other related measures of COVID-19 in the UK}
\author{T.~Maishman$^\dagger$, S.~Schaap$^\dagger$, D.~S.~Silk$^\dagger$, S.~J.~Nevitt$^\ddagger$, D.~C.~Woods$^{\dagger\dagger}$, and V.~E.~Bowman$^\dagger$ \\
\small $^\dagger$Defence Science and Technology Laboratory, Porton Down, UK \\
\small $^\ddagger$Department of Biostatistics, University of Liverpool, UK \\
\small $^{\dagger\dagger}$Statistical Sciences Research Institute, University of Southampton, UK}

\begin{document}

\maketitle

\thispagestyle{fancy}

In the recent COVID-19 pandemic, a wide range of epidemiological modelling approaches have been used to predict the effective reproduction number, $R(t)$, and other COVID-19 related measures such as  the daily rate of exponential growth, $r(t)$. These candidate models use different modelling approaches or differing assumptions about spatial or age mixing, and some capture genuine uncertainty in scientific understanding of disease dynamics. Combining estimates using appropriate statistical methodology from multiple candidate models is important to better understand the variation of these outcome measures to help inform decision making.
In this paper, we combine these estimates for specific UK nations and regions using random effects meta analyses techniques, utilising the restricted maximum likelihood (REML) method to estimate the heterogeneity variance parameter, and two approaches to calculate the confidence interval for the combined estimate: the standard Wald-type intervals; and the Knapp and Hartung (KNHA) method. 
As estimates in this setting are derived using model predictions, each with varying degrees of uncertainty, equal weighting is favoured over the more standard inverse-variance weighting in order avoid potential up-weighting of models providing estimates with lower levels of uncertainty that are not fully accounting for inherent uncertainties. 
Both equally weighted models using REML alone and REML+KNHA approaches were found to provide similar variation for $R(t)$ and $r(t)$, with both approaches providing wider, and therefore more conservative, confidence intervals around the combined estimate compared to the standard inverse-variance weighting  approach.  
Utilising these meta-analysis techniques has allowed for statistically robust combined estimates to be calculated for key COVID-19 outcome measures.  This in turn allows timely and informed decision making based on all of the available information.

\section{Introduction}

Following the outbreak of COVID-19 and attempts to control the spread of the disease, focus in the UK has moved to estimating the effective reproduction number, $R(t)$, which reflects the infectious potential of a disease and is defined as the average number of secondary cases per primary case at time $t$ since the start of the epidemic\cite{Wallinga2004}. The basic reproduction number, $R(0)$, is the number of secondary cases per primary case at the beginning of an epidemic, in an entirely susceptible population\cite{RoyalSoc2020}. As more individuals are infected or immunised, the population in which $R(t)$ is based consists of both naive/susceptible and exposed/immune individuals and therefore changes over time\cite{RoyalSoc2020}. If $R(t)$ for the UK exceeds 1, the infection rate will grow exponentially. To bring the epidemic under control, the corresponding $R(t)$ needs to drop and remain as far below 1 as practicable\cite{Wallinga2004}. There are a number of ways to estimate $R(t)$, for example using information on the number of cases, number of deaths, survey data, or a combination of these. From incidence/cases data, the mean generation time and initial growth rates (defined as the \textit{per capita} change in number of new cases per unit of time) in the infected population can be used\cite{Wallinga2004, Wallinga2007}. From death data, $R(t)$ can be determined by using the number of deaths that can be attributable to the infection, with key information including the infection fatality rate, mean generation time and the time from onset of symptoms to death\cite{Flaxman2020, Ridenhour2014}. For example, $R(t)$ can be linked to the number of deaths using a renewal equation which incorporates the time between death of the infector and infectee\cite{RoyalSoc2020}. $R(t)$ can also be determined by surveying the population for infection and inferring likely case data; an approach  which commonly uses a contact function that identifies the susceptible individuals, how likely transmission is to be (given that contact has occurred), and measures the contact between members of the population \cite{Kucharski2020,Farrington2001}. Detailed methodology is not provided in this paper but available from the Royal Society\cite{RoyalSoc2020}. 

Other key COVID-19 outcomes of interest include the daily rate of exponential growth, $r(t)$, which represents an approximation of the percentage change in the number of infections over time\cite{DHSC2020}. If $r(t)$ is positive, the infection rate will grow exponentially, whereas if $r(t)$ is negative and remains negative, it will be possible for the epidemic to be brought under control. 

In the UK, epidemiological modelling is provided by a number of highly skilled academic groups based on a number of different data streams, modelling techniques and assumptions (a summary of these models is provided in the appendix and detailed descriptions are also available from the Royal Society\cite{RoyalSoc2020}).  Each of these groups provide key understanding and insight into the current state of the epidemic, and these estimates must therefore be combined to provide an overall assessment so that decision making is based on all available evidence. In this paper, we use meta-analyses to combine estimates of $R(t)$ and $r(t)$ for specific nations/geographical regions of the UK, from multiple candidate epidemiological models.  

\subsection{Existing Methods to Combine Estimates}

The methodology used to combine modelling estimates is not limited to meta analyses. For example, Lindstrom \textit{et al} incorporate an ensemble modelling approach using a Bayesian framework and various weighting schemes\cite{Lindstrom2015}. Ensemble methods were also explored by Ray \textit{et al}\cite{Ray2018}, which used model stacking\cite{Wolpert1992}, again, with exploration into different weighting approaches to combine predictions from multiple models\cite{Ray2018}. Methods used to aggregate expert-generated predictions have also been explored by Genest and Zidek, O'Hagan \textit{et al}, and McAndrew \textit{et al}\cite{Genest1986, OHagan2006, McAndrew2021}. Genest and Zidek provide a comprehensive annotated bibliography on various methods, including but not limited to: the use of a supra Bayesian approach whereby in some cases, there is a decision maker for whom the panel of experts reports to\cite{Ratcliff1979, Thomas1980}; and the Vincentization method which averages the per cent quantiles of the experts' distribution to construct a consensus distribution\cite{Keeney1976}. McAndrew \textit{et al} provide a more recent review on various methods to aggregate predictions from experts, including Cooke's method which incorporates a calibration score to assign weights to the experts\cite{Cooke1991}, stacking methods\cite{Wolpert1992}, and other pooling methods which transform the aggregated forecast distribution such as the Spread-adjusted Linear Pool (SLP) method\cite{Berrocal2007, Glahn2009, Kleiber2011} and Beta Linear Pool (BLP) method\cite{Ranjan2010, Gneiting2013}. In terms of combining COVID-19 related outcomes, a number of combination approaches were explored to combine model projections by Silk \textit{et al}\cite{Silk2020} and Funk \textit{at al}\cite{Funk2020}, including stacking methods, and regression-based methods such as ensemble model output statistics (EMOS)\cite{Gneiting2005}, and Quantile Regression Averaging (QRA)\cite{Nowotarski2015}. 

\subsection{Application of the Meta Analysis Approach}

Meta analysis, the process of synthesizing data from a series of separate studies\cite{Borenstein2009}, is a well-known and established method, used ubiquitously in fields such as epidemiology, medicine, climate science, psychology, and education. It provides a rapid and simple approach, and its results are easy to interpret. In this paper, we use this method to provide an estimate of $R(t)$ from multiple models and assumptions.  Effectively $R(t)$ is a physical quantity that could potentially be measured if we had perfect knowledge of infection state and transmission risk of all individuals through time.  Clearly, in reality, this is impossible and therefore $R(t)$ must be estimated from available data.  However, there are a number of entirely valid ways to estimate $R(t)$ and each provides insight into the current value. We require the best knowledge of $R(t)$ that can be provided and each model estimate captures an aspect of the current $R(t)$ value, therefore meta-analysis will, by definition, provide an overall estimate, averaged over all of the modelling assumptions and potential methodologies, providing a combined estimate that benefits from all available information. However, the combination naturally assumes that the candidate models are valid and worth considering.

Meta-analysis models can assume fixed or random effects; i.e. a shared common effect or distribution of effects. As it is possible for each candidate model to use a different method to estimate these outcome measures, the modelling approaches and/or underlying assumptions are assumed to vary. For example, different modelling approaches (e.g. mechanistic or empirical) or differing assumptions about spatial or age mixing may be used\cite{Silk2020}. Moreover, the random effects model assumes a distribution of true effect sizes as opposed to a shared common (true) effect size assumed in the fixed effects model\cite{Borenstein2009, Langan2019}. Subsequently, a meta-analysis using a random effects model is chosen over a fixed effects model. Details and motivating examples on fixed and random effects models for analysis can be found in Borenstein \textit{et al.}\cite{Borenstein2010}. The random effects model can be defined as:
\begin{equation}\label{random_effects_model}
\widehat{\theta_i}=\theta_i+\varepsilon_i \;\;\;\;\;\;\;\; \theta_i\sim{N}(\theta, \tau^2)\,
\end{equation}
where $\theta_i$ is the true effect size in group $i$ (for a set of $i=1, ..., k$ groups), $\widehat{\theta_i}$ is the estimated effect size in group $i$, $\theta$ is the average effect across all groups, and $\varepsilon_i$ are the within‐group errors\cite{Langan2019}. $\theta_i$ is sampled from a distribution, typically assumed to be normal, of mean $\theta$ and variance $\tau^2$, the heterogeneity variance parameter.\cite{Langan2019}. 

The combined estimate, $\widehat{\theta}$, with associated variance, $Var(\widehat{\theta})$, can be calculated as follows\cite{Langan2019}:

\begin{equation}\label{theta_var}
\begin{gathered}
\widehat{\theta}=\frac{\sum\limits_{i=1}^k w_i \widehat{\theta_i}}{\sum\limits_{i=1}^k w_i}, \\
Var(\widehat{\theta})=\left(\frac{1}{\sum\limits_{i=1}^k w_i}\right)^2 \sum\limits_{i=1}^k {w_i}^2 (\widehat{\sigma_i}^2 + \widehat{\tau}^2)
\end{gathered}
\end{equation}
where $w_i$ denotes the weighting applied to the estimate in group $i$, $\widehat{\sigma_i}^2$ the estimated variance of the estimate in group $i$, and $\widehat{\tau}^{2}$ the estimated heterogeneity variance parameter; a measure of the heterogeneity (or variability) between estimates.

The standard weighting applied in a meta analysis is by way of \textit{inverse-variance}, whereby $w_i=1/(\widehat{\sigma_i}^2 + \tau^2)$, whereas an equally weighted model has weighting $w_i=1/k$. The corresponding combined estimate, $\widehat{\theta}$, and associated variance, $Var(\widehat{\theta})$, from Equation \eqref{theta_var} become:


\begin{equation*}
\hat{\theta} \quad = \quad
\begin{cases}
\frac{\sum\limits_{i=1}^k \widehat{\theta_i}(\widehat{\sigma_i}^2 + \widehat{\tau}^2)^{-1}}{\sum\limits_{i=1}^k (\widehat{\sigma_i}^2 + \widehat{\tau}^2)^{-1}} & \text{for inverse-variance weighting} \\
\frac{1}{k}\sum\limits_{i=1}^k \widehat{\theta_i}& \text{for equal weighting}
\end{cases}\\
\end{equation*}

\begin{equation}\label{theta_var_with_weights}
Var(\widehat{\theta}) = 
\begin{cases}
 \frac{1}{\sum\limits_{i=1}^k (\widehat{\sigma_i}^2 + \widehat{\tau}^2)^{-1}} & \text{for inverse-variance weighting} \\
 \frac{1}{k^2}\sum\limits_{i=1}^k (\widehat{\sigma_i}^2 + \widehat{\tau}^2)& \text{for equal weighting}
\end{cases}
\end{equation}

For random effects meta analyses, several methods are available to estimate $\tau^2$. In addition, multiple methods can be used to calculate the confidence intervals (CIs) for the combined estimate. This paper focuses on the well-established restricted maximum likelihood (REML) method recommended by Veroniki \textit{et al.}\cite{Veroniki2016} to estimate $\tau^2$, with the incorporation of two different approaches for the calculation of the CIs: the standard Wald-type method; and the Knapp and Hartung (KNHA) method (also referred to as the Hartung-Knapp-Sidik-Jonkman method)\cite{Hartung2001, Sidik2002}. The Wald-type method is chosen as it is a well-established approach, whilst the KNHA method has been shown to provide better coverage\cite{Langan2019}. The standard Wald-type CI is calculated as\cite{Langan2019}:

\begin{equation}\label{wald}
\widehat{\theta}{\pm}z_{1-\frac{\alpha}{2}}\sqrt{Var(\widehat{\theta})} 
\end{equation}
with $\widehat{\sigma_i}^{2}$ the estimated variance for group $i$, and $z$-score calculated for the required confidence interval of the standard normal distribution.

The KNHA CI is calculated as\cite{Langan2019}:
\begin{equation}\label{knha}
\begin{gathered}
\widehat{\theta}{\pm}t_{k-1,1-\frac{\alpha}{2}}\sqrt{Q \cdot Var(\widehat{\theta})} \\
\mbox{where} \;\; Q = \frac{1}{k-1}\sum\limits_{i=1}^k \left(\frac{1}{\widehat{\sigma_i}^2 + \widehat{\tau}^2}\right)(\widehat{\theta_i}-\widehat{\theta})^2
\end{gathered}
\end{equation}
with $t$-score calculated from the $t$ distribution with $k-1$ degrees of freedom.

The use of REML to estimate $\tau^2$ has been shown to be robust to deviations from normality and to perform well, particularly when utilising the KNHA method to calculate the CIs, when only a limited number of models are available for comparison\cite{Kontopantelis2012, Kontopantelis2012a, Langan2019}. This papers refers to these two approaches as the \textit{REML alone} and \textit{REML+KNHA} approaches respectively. 

\section{Methods}

\subsection{Data Preparation}

This paper utilised data from 12 different candidate models, in which estimated quantiles from each model were available for up to 12 UK nations/regions for a set cut-off date. These candidate models were drawn from many of the leading academic institutions and epidemiologists in the UK whose models already support government response for pandemics.  In this paper, candidate models and UK nations/regions were anonymised, and estimates were combined according to each of the anonymised UK nations/regions separately.

The aim of the data preparation step is to generate appropriate estimated means and standard errors for each candidate model to be used in the combination. For a set of $i=1, ..., k$ candidate models, let $y_i$ denote the mean estimate of the outcome measure of interest for the $i^{th}$ model (previously denoted $\theta_i$), with associated standard error, $se_i$.  

Each of the candidate models outputs $j^{th}$ percentiles, $Q_{i}(j)$, for the outcome measure of interest, as opposed to $y_i$ and $se_i$. In order for the estimates to be combined in a random effects model for an outcome measure of interest, initial approximations of $y_i$ and $se_i$, $\widehat{y_i}$ and $\widehat{se_i}$ are required. Using the $j^{th}$ percentiles from the $i^{th}$ candidate model, $Q_{i}(j)$, $y_i^*$ and $se_i^*$ are initially calculated as follows:
\begin{equation}\label{yi}	
y_i^* = Q_i(50)\,
\end{equation} \\[1pt]
\begin{equation}\label{standard_error}
se_i^* = \frac{max( |Q_i(95) - Q_i(50)|, |Q_i(50) - Q_i(5)|)}{z_{1-\frac{\alpha}{2}}}\,
\end{equation}
with $z$-score calculated using $\alpha=0.10$ for the 90\% confidence interval of the standard normal distribution.

\subsection{Skewness Exploration and Correction}

As some of the model estimates maybe skewed, the use of $Q_i(50)$ for an approximation of $yi$ may not be optimal and an adjusted estimate required. First, the degree of skewness of the estimates, $SK_i$, is calculated and assessed using Bowley's formula\cite{Bowley1920}:
\begin{equation}\label{skewness}
SK_i = \frac{Q_i(75) + Q_i(25) - 2Q_i(50)}{Q_i(75) - Q_i(25)}\,
\end{equation}

An absolute value of 0.5 is then used to indicate a moderate or higher level of skewness\cite{Bulmer1979}. If $|SK_i|\leq0.5$, then skewness is deemed sufficiently small and a normal distribution can be fitted to the percentiles, i.e. $\widehat{y_i}=Q_i(50)$ from Equation \eqref{yi} and $\widehat{se_i}=se_i^*$ from Equation \eqref{standard_error}. However, if $|SK_i|>0.5$, then an adjustment to the estimates are required. First, appropriate transformations to the percentiles are made: if the estimates are negatively skewed the quantiles are inverted i.e. $Q_i(5)$, $Q_i(50)$, $Q_i(95)$  $\,\to\,$  -$Q_i(95)$, -$Q_i(50)$, -$Q_i(5)$; and a positive constant is added, where applicable, to ensure the adjusted quantiles are positive. A gamma distribution is fit to the adjusted percentiles by minimizing the sum of squared distance between the percentiles of the gamma distribution and those of the model estimates using a Particle Swarm Optimisation (PSO) algorithm\cite{Kennedy1995}.  The PSO is performed using the \texttt{psoptim} optimisation call from the \texttt{pso} package\cite{Bendtsen2012} in R\cite{RCoreTeam2019}. This optimizes the non-linear function via an algorithm using a series of learning parameters\cite{Kennedy1995}. Further details on the process are provided by Kennedy\cite{Kennedy1995}, Yang\cite{YANG201499}, and Bendtsen\cite{Bendtsen2012}. The square-root of the variance from the optimisation process can then be used as a conservative estimate of $\widehat{se_i}$, and the corresponding mean from the optimisation process, after a suitable back-transformation applied, can be used for $\widehat{y_i}$.  Although the adjusted estimates remain skewed, the use of REML for a meta-analysis is robust even in the case of extreme non‐normal distributions\cite{Kontopantelis2012, Kontopantelis2012a}.

\subsection{Equal Weighting}

The standard weighting applied in meta analyses is by way of \textit{inverse-variance} weighting, whereby estimates which provide the highest precision are weighted highest. However, estimates in this setting are derived using model predictions, each with varying degrees of uncertainty, i.e. estimates provided with smaller levels of uncertainty are not necessarily more representative of the situation over another model. For example, a model with wider 90\% intervals could in fact be more representative over another model with narrower 90\% intervals as the modelling approach takes into account more information in the derivation of its estimates. The standard \textit{inverse-variance} weighting could therefore unjustifiably change estimates as models with smaller uncertainty will be up-weighted. As each modelling approach differs in how uncertainty is accounted for and conservative estimates in the context are preferable, the comparison of uncertainty levels alone would not be appropriate in this particular setting. To counter this, user-defined equal weighting is applied to the candidate models using $\frac{1}{k}$, where $k$ is the number of candidate models that are included in the random effects model\cite{Bonett2008}.  

\subsection{Fitting the Random Effects Model}

Having estimated the distributions of each model to be included in the combination, we now calculate the combined estimate using the random effects model.
The custom weights, together with $\widehat{y_i}$ and $\widehat{se_i}$ from the fitted distributions of each candidate model, are passed to the \texttt{metafor} package in R using the \texttt{rma} call\cite{Viechtbauer2010}, using the REML method to estimate $\tau^2$ with incorporation of either the Wald-type CIs (\textit{REML alone}), or KNHA method for the calculation of the CIs (\textit{REML+KNHA}). 

\section{Worked Example}\label{example}

To illustrate the method in practice, a step by step guide is given here for how the estimated quantiles from a group of anonymised models for a selected anonymised UK nation/region can be used to provide a combined estimate for this selected nation/region. A full set of results for all UK nations/regions can be found in Section 5 and the Appendix, and a csv file and example R script provided as supplementary material for the worked example. 
Table \ref{example:r_ex_data} shows the $R(t)$ estimated quantiles from 12 anonymised models for anonymised UK nation/region 10, together with the calculated $se_i^*$ and $SK_i$ using Equations \eqref{standard_error} and \eqref{skewness} respectively, and corresponding $\widehat{y_i}$ and $\widehat{se_i}$ calculated values. No estimated quantiles were available from candidate model 8 for this particular nation/region but estimated quantiles are available for other nation/regions for this model (see Appendix Table \ref{appendix:r_data} for the full list of $R(t)$ estimates by model and nation/region). 

Moderate to high skewness was identified for candidate model 5, although this was only marginal ($8.6 \times 10^{-14}$ over the threshold). The corresponding adjusted estimate, $\widehat{y_i}$, following input into the \texttt{psoptim} optimisation call resulted in an identical estimate to $Q_i(50)$ in this case (to four decimal places), but with modified $\widehat{se_i}$ of 0.0028.

	\begin{table}[!htb] 
      \centering
	\begin{tabular}{@{}cccccccccc@{}}
		\toprule
			Model & $Q_i(5)$ & $Q_i(25)$ & $Q_i(50)$ & $Q_i(75)$ & $Q_i(95)$ & $SK_i$ & $se_i^*$ & $\widehat{y_i}$ & $\widehat{se_i}$ \\ \midrule
			1  & 0.6300 & 0.6800 & 0.7400 & 0.8100 & 0.8700 & 0.0769 & 0.0790 & 0.7400 & 0.0790 \\
			2  & 0.6228 & 0.6775 & 0.7045 & 0.7413 & 0.8265 & 0.1540 & 0.0742 & 0.7045 & 0.0742 \\
			3  & 0.6400 & 0.7000 & 0.7400 & 0.7900 & 0.8700 & 0.1111 & 0.0790 & 0.7400 & 0.0790 \\
			4  & 0.4400 & 0.6300 & 0.7500 & 0.8700 & 1.1400 & 0.0000 & 0.2371 & 0.7500 & 0.2371 \\
			5  & 0.7898 & 0.7930 & 0.7954 & 0.7963 & 0.7995 & -0.5000 & 0.0034 & 0.7954 & 0.0028 \\
			6  & 0.8076 & 0.8199 & 0.8329 & 0.8494 & 0.8749 & 0.1189 & 0.0256 & 0.8329 & 0.0256 \\
			7  & 0.6232 & 0.7111 & 0.7862 & 0.8647 & 0.9890 & 0.0222 & 0.1233 & 0.7862 & 0.1233 \\
			8  & - & - & - & - & - & - & - & - & - \\
			9  & 0.7509 & 0.8626 & 0.9382 & 1.0159 & 1.1604 & 0.0148 & 0.1351 & 0.9382 & 0.1351 \\
			10  & 0.8175 & 0.8250 & 0.8302 & 0.8353 & 0.8427 & -0.0041 & 0.0077 & 0.8302 & 0.0077 \\
			11  & 0.8412 & 0.8956 & 0.9293 & 0.9657 & 1.0340 & 0.0398 & 0.0637 & 0.9293 & 0.0637 \\
			12  & 0.6600 & 0.7100 & 0.7600 & 0.8000 & 0.8600 & -0.1111 & 0.0608 & 0.7600 & 0.0608 \\ \bottomrule
	\end{tabular}
	\caption{$R(t)$ estimates and corresponding $SK_i$, $se_i^*$, $\widehat{y_i}$ and $\widehat{se_i}$ calculated values for anonymised models 1 to 12 for anonymised UK nation/region 10. All numbers displayed to four decimal places. $^\dagger$No estimated quantiles were available from candidate model 8 for this particular nation/region. }
	\label{example:r_ex_data}
	\end{table}

\begin{figure}[H]
	\centering
		\includegraphics[width=1\textwidth]{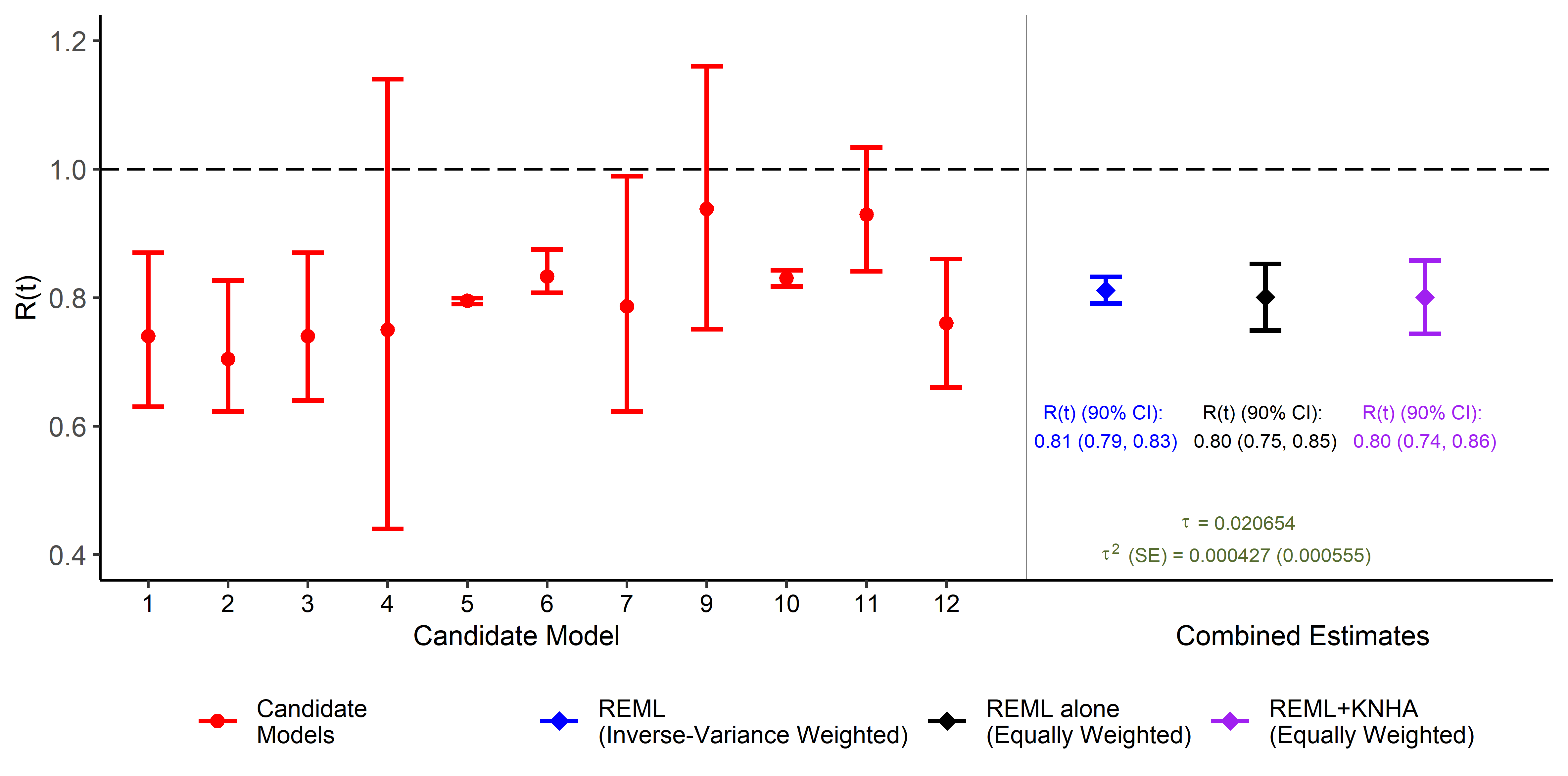}
	\caption{$R(t)$ estimates from the candidate models for anonymised nation/region 10, together with calculated combined estimates using: an \textit{inverse-variance} weighted approach with Wald-type CIs; an equally weighted approach with Wald-type CIs (\textit{REML alone}); and an equally weighted approach with KNHA CIs (\textit{REML+KNHA}). The error bars illustrate the 90\% CIs.}
	\label{example:r_ex_figure}
\end{figure}

To illustrate the performance of the equal weighting random effects model approach, an initial random effects model using the REML method to estimate $\tau^2$ but with the standard \textit{inverse-variance} weighting was applied to provide a combined estimate. The equally weighted random effects models using REML and Wald-type CIs (\textit{REML alone}) or KNHA CIs (\textit{REML+KNHA}) were then applied to the same estimates. The $R(t)$ estimates from the candidate models, together with the combined estimates using these methods are shown in Figure \ref{example:r_ex_figure}.

The combined estimate obtained is 0.81 for \textit{inverse-variance} weighted approach, and 0.80 for each of the equally weighted approaches, with 90\% CIs ranging from 0.79 to 0.86 indicating that we can be reasonably sure the true $R(t)$ for this particular region at time $t$ is below 1.
As mentioned above, estimates in this setting are derived using model predictions, and a model with wider 90\% intervals could in fact be more representative of the situation when there is inherent uncertainty throughout multiple data collection and modelling streams than a model with narrower 90\% intervals using fewer data streams. The results shown in Figure \ref{example:r_ex_figure} show that the \textit{inverse-variance} weighted approach produced narrower 90\% CIs compared to either of the equally weighted approaches. As $\tau^2$ is very small, the standard error of the estimate dominates the inverse variance weighting, and so this narrow 90\% interval is primarily driven by the estimates from candidate models 5 and 10, which had narrower 90\% intervals compared to the other candidate models.  Conversely, candidate model 4 contributed little information to the combined \textit{inverse-variance} weighted estimate due to the wider 90\% intervals provided. This example highlights a key advantage of the equally weighted approach in this particular setting; the ability to avoid potential up-weighting of models providing estimates with lower levels of uncertainty that are not fully accounting for inherent uncertainties. Both the \textit{REML alone} and \textit{REML+KNHA} equally weighted approaches provided similar results in this worked example. However, a more in-depth look at the differences between the results obtained from these two methods is explored in the Results section, below. 

\section{Results}\label{results}

A full set of results for $R(t)$ and $r(t)$ for the 12 anonymised candidate models is provided across 12 anonymised UK nations/regions below. The estimate for $\tau^2$ for each outcome measure and region is provided in the Appendix. 

\subsection{Combined \textit{R(t)} Estimates}

\begin{figure}[H]
	\centering
		\includegraphics[width=1\textwidth]{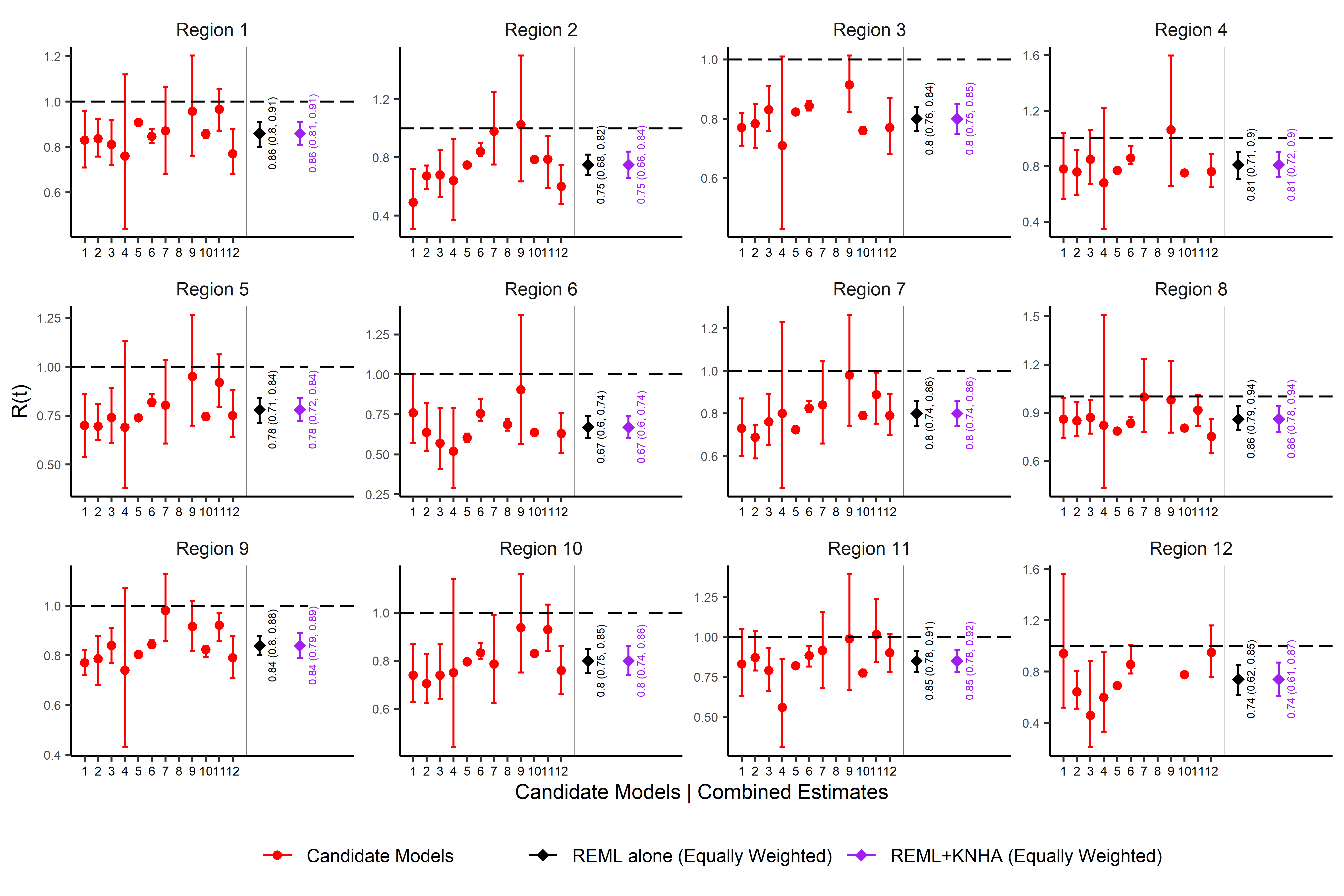}
	\caption{$R(t)$ estimates from the candidate models by anonymised nation/regions, together with calculated combined estimates using equally weighted models, with \textit{REML alone} or \textit{REML+KNHA} approaches for the 90\% CIs. The error bars illustrate the 90\% CIs.}
	\label{results:r_figure}
\end{figure}

The $R(t)$ estimates by region for the candidate models are shown in Figure \ref{results:r_figure}. The upper 90\% CIs were lower than 1 for all individual regions indicating that that we can be reasonably sure that $R(t)$ for all individuals regions at time $t$ was below 1. On visual inspection, the difference in 90\% CI for $R(t)$ between equally weighted models using \textit{REML alone} versus \textit{REML+KNHA} approaches was minimal. On closer inspection of the combined estimates to additional decimal places (data not shown), in seven of the 12 regions the \textit{REML+KNHA} approach provided a wider and more conservative 90\% CI than the \textit{REML alone} approach, compared to five instances where the \textit{REML alone} approach provided a wider 90\% CI than the \textit{REML+KNHA} approach. Looking at models across different regions, candidate model 4 consistently had wider 90\% intervals compared to the other candidate models, whilst candidate models 5 and 10 consistently had narrower 90\% intervals. The $\tau^2$ estimates for all regions were again very small (see Table \ref{appendix:r_data} in the Appendix), indicating that the standard error of the estimate dominates the inverse variance weighting, which, coupled with the large disparity in uncertainty for estimates in each region, highlights the appropriateness of applying equal weighting to the models in this setting. Moreover, the equal weighted approaches provided wider 90\% CIs compared to the \textit{inverse-variance} weighting approach for all regions (Table \ref{appendix:r_data}).

\subsection{Combined \textit{r} Estimates}

\begin{figure}[H]
	\centering
		\includegraphics[width=1\textwidth]{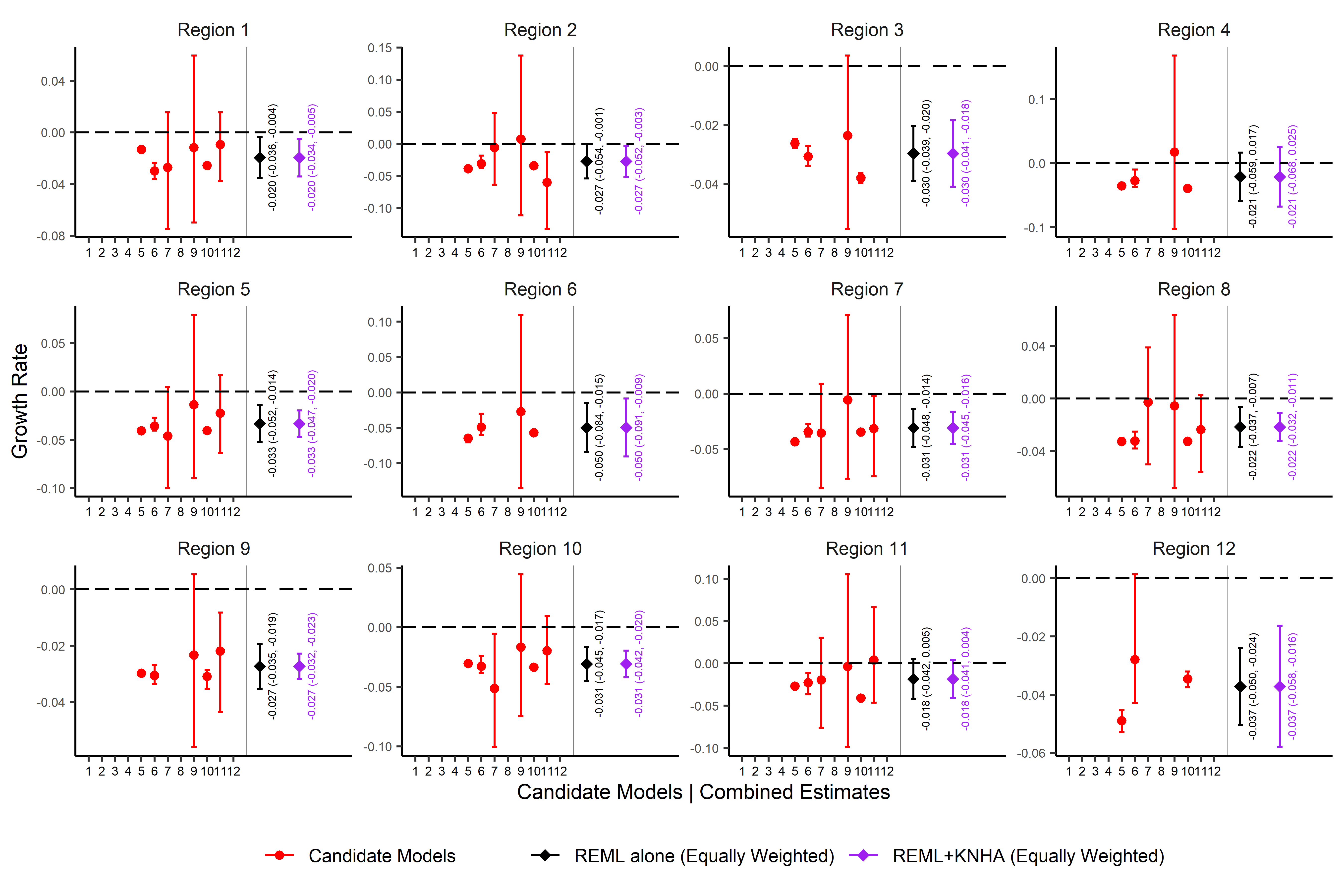}
	\caption{$r(t)$ estimates from the candidate models by anonymised nation/regions, together with calculated combined estimates using equally weighted models, with \textit{REML alone} or \textit{REML+KNHA} approaches for the 90\% CIs. The error bars illustrate the 90\% CIs.}
	\label{results:gr_figure}
\end{figure}

In terms of $r(t)$ (Figure \ref{results:gr_figure}), initial visual inspection yielded a similar conclusion to the combined estimates for $R(t)$. The 90\% CIs were equal to or lower than zero for all individual regions indicating that that we can be reasonably sure that $r(t)$ for all individuals regions was not increasing. Only slight differences were found in the 90\% CI estimates between the two approaches. However, in this case, closer inspection of the estimates indicated that in eight of the 12 regions the \textit{REML alone} approach provided a wider 90\% CI than the \textit{REML+KNHA} approach, compared to four instances where the \textit{REML+KNHA} approach provided a wider 90\% CI than the \textit{REML alone} approach. Looking at models across regions, it is first important to note that there were only half of the candidate models for which estimates were available for $r(t)$ compared to estimates for $R(t)$, particularly evident for region 12, in which only three candidate models were included. In terms of variability, candidate models 5 and 10 once again consistently had narrower 90\% intervals across regions, whilst candidate model 9 consistently had wider 90\% intervals. Although the $\tau^2$ estimates for all regions were again small for $r(t)$, showing low inter-model variability, the equally weighted approaches provided moderately wider 90\% CIs compared to the \textit{inverse-variance} weighting approach for all regions (see Table \ref{appendix:gr_data} in the Appendix), which is preferable where there is the potential that uncertainty is arising outside of the scope of some modelling approaches.

\section{Discussion}

When comparing the results of the \textit{REML alone} and \textit{REML+KNHA} approaches, both provided almost identical results for $R(t)$, and very similar results for $r(t)$. In addition, both approaches provided more conservative CIs around the combined estimate compared to the standard \textit{inverse-variance} weighting  approach.  

\begin{figure}[H]
	\centering
		\includegraphics[width=1\textwidth]{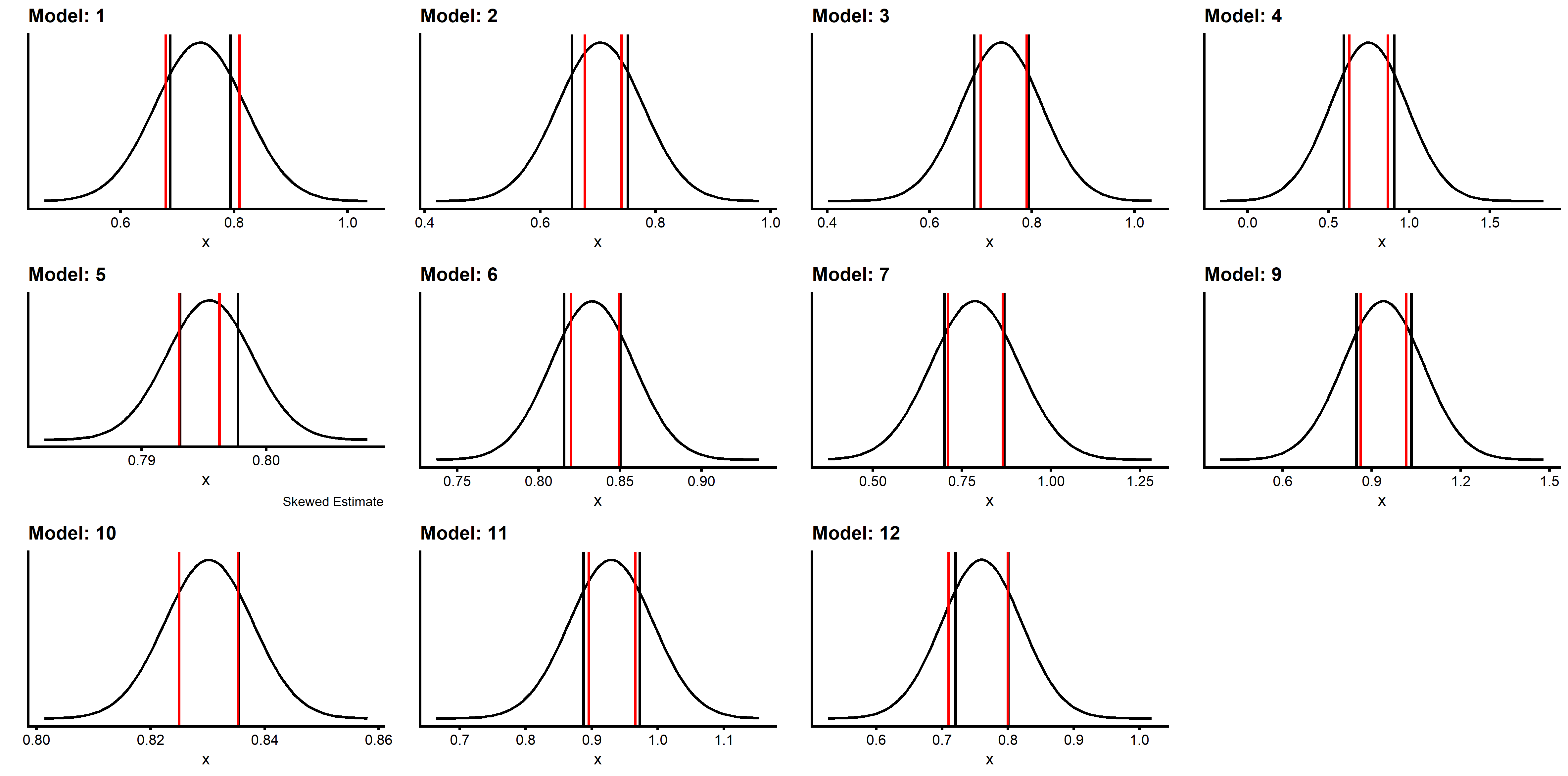}
	\caption{Normal distributions generated using mean of $y_i^*$ and standard deviation of $se_i^*$ from Equations \eqref{yi} and \eqref{standard_error} for $R(t)$ estimates from the candidate models for anonymised nation/region 10. Black vertical lines represent the $25^{th}$ and $75^{th}$ percentiles drawn from the generated normal distributions whilst the red vertical lines illustrate the $25^{th}$ and $75^{th}$ percentiles obtained directly from the candidate models. The plot for Model 5 is marked as skewed as per the result obtained using the skewness calculation in Equation \eqref{skewness}.}
	\label{appendix:quantile_assessment}
\end{figure}

There are a number of possible extensions to the methodology presented. For example, $y_i$ are assumed to be unbiased and normally-distributed estimates of the corresponding true effect\cite{Viechtbauer2010}, and alternative approaches to approximate $y_i$ and $se_i$ may be used. However, as noted in the Cochrane Handbook for Systematic Reviews of Interventions, a median will be very similar to the mean when the distribution of the data is symmetrical\cite{Higgins2019}. Moreover, the use of the square-root of the variance from the optimisation process to approximate $se_i$ enables a larger estimate to be provided, and thus a more conservative degree of uncertainty. Alternative methods to calculate the standard deviation (to be used for an approximation of $se_i$) such as those outlined by Bland\cite{Bland2015} and Wan \textit{et al}\cite{Wan2014} are not possible due to the lack of availability of the sample size, minimum and maximum in this setting. Wan \textit{et al}\cite{Wan2014} notes the use of $Q_i(75)-Q_i(25)/1.35$ taken from the Cochrane Handbook\cite{Higgins2019}, however as noted in the Cochrane Handbook, this approximation is for instances with large sample sizes.  In addition, Figure \ref{appendix:quantile_assessment} shows the normal distributions generated using mean of $y_i^*$ and standard deviation of $se_i^*$ from Equations \eqref{yi} and \eqref{standard_error} for $R(t)$ estimates from the candidate models for anonymised nation/region 10. This provides a visual confirmation of the fit of the candidate model percentiles against the drawn distributions (with the exception of Model 5 which is marked as skewed as per the result obtained using the skewness calculation in Equation \eqref{skewness}). Finally, and of key importance, it has been shown that the performance of statistical methods, such as REML for a meta-analysis, are robust, even in the case of extreme non‐normal distributions\cite{Kontopantelis2012, Kontopantelis2012a}. 

It should be noted that some models rely on similar data streams for their primary information, and there is likely a spatial relationship between regional estimates from the same group. In terms of similar data streams, the model structures are all different, and a large amount of variation is observed in the estimates. Consequently, the impact on the results is extremely limited. To illustrate this degree of impact, a sensitivity analyses on the $R(t)$ estimates was performed using the \texttt{rma.mv} call from the \texttt{metafor} package\cite{Viechtbauer2010}, which enables a model to be fitted for dependent effect sizes. An equally weighted model, using REML and Wald-type CIs, was formulated with model number fitted as the inner-most random effect, and data type fitted as the outer-most random effect in the model. The results were almost identical to the univariate equally weighted model (using REML and Wald-type CIs), with no differences observed larger than 0.01. It should be noted that at the time of writing, the \texttt{rma.mv} call does not have the ability to incorporate the \textit{REML+KNHA} approach and so this comparison was not possible. In terms of any dependence between regional estimates from the same group, any correlation assumptions are not consistent between models and as a result, this is outside of the scope of this paper. However, the authors acknowledge that future work in this area might be worth exploring. A final remark in terms of possible correlations between the metrics of interest should also be made here. However, although $R(t)$ and $r(t)$ are probably correlated, not all groups provide both sets of estimates for these, and more importantly, not all candidate models are modelled in the same way between groups and the degree in which $R(t)$ and $r(t)$ are correlated will vary i.e., they may have differing correlation structures, etc. As a result, it is not possible to accurately carry this out without making further untestable assumptions regarding the different correlation structures. 

The assumption that all candidate models are valid/plausible is important to note, however each model uses different ways to estimate $R(t)$, which are all equally valid and each provides insight into the current value. Inclusion of a variety of approaches is crucially important as any subgroup of models could lead to potential up-weighting of models providing estimates with lower levels of uncertainty that are not fully accounting for inherent uncertainties. For these reasons, the incorporation of equal weighting has been chosen. The use of equal weighting in meta analyses is not novel and as noted by Borenstein \textit{et al}\cite{Borenstein2009}, its application has actually been recommended in some papers \cite{Bonett2008, Bonett2009, Shuster2010}. The purpose of this paper is not to advocate the use of the approach in general meta analyses settings, but for this particular setting. It is also important to note that $R(t)$ is in effect impossible to measure as it would require perfect knowledge of all individuals through time, and there are therefore no 'gold-standards' to compare the individual (and combined) estimates to. There are, however, real world assessments of these data which align, but have potential natural sampling bias (and are therefore not a gold standard), for example: the Office for National Statistics (ONS) survey which covers estimates for England, Wales, Scotland and Northern Ireland\cite{ONS2020}; the CoMix study which consists of a survey of UK adults\cite{Jarvis2020}; and the REACT (REal-time Assessment of Community Transmission) study which incorporates a series of studies using home testing on people across England\cite{REACT2021}. When the model estimates are combined therefore, and despite potential natural sampling bias, informal comparisons can be made against these survey estimates to help provide approximate feasibility checks on the results.

The authors also acknowledge that, whilst meta analyses in this setting was chosen as it is well established and able to provide rapid results which are easy to interpret,  there are other methods that could be applied to combine estimates, such as various ensemble modelling approaches\cite{Lindstrom2015, Ray2018, Wolpert1992}, expert elicitation\cite{OHagan2006}, the use of a supra Bayesian approach\cite{Ratcliff1979, Thomas1980}. 

Another possible extension is in regards to the use of combining estimates for an entire region, i.e. not splitting the regions into urban versus rural areas, or not taking into account the number of care homes, etc. Indeed, by definition $R(t)$ is an average over a population.  However, if the population in question is very heterogeneous in space or the models used to estimate $R(t)$ become unreliable due to very low case numbers (in this situation case numbers are stochastic and not well approximated by exponential models) then $R(t)$ may not be an appropriate measure. However, in order to address this and ensure that any combination is representative, a basic reliability score is also calculated for use when interpreting these results for a specific region. The reliability score uses estimated case numbers in the modelled region and the heterogeneity in space of the numbers of cases (e.g. a dense urban outbreak compared to rural areas with no cases)\cite{DHSC2021}. It should also be noted that each model provides estimates for each region individually, i.e. estimates for all English regions were not combined to get an overall estimate for England. The use of a reliability score for each region when presenting the results enables for a more measured conclusion to be drawn from the combined estimates for each region. Further investigation into the reliability score and combining estimates for smaller spatial regions is likely to form part of future work in this area.  

Finally, many of the candidate models provide estimates of $R(t)$ over specific time periods, thus providing estimates of $R(t)$ at a specific date. We would therefore like to explore predicting $R(t)$ as a time series, as opposed to at a specific time point, which is particularly important if $R(t)$ changes rapidly over time. Further to this, we would like to explore predicting the probably that $R(t)$ is changing and how rapidly it is changing, using historical combined estimates of $R(t)$ as a prior.

\section{Summary}

This paper describes appropriate statistical methodology to provide a combined estimate of effective reproduction number, $R(t)$, and the daily rate of exponential growth, $r(t)$, of COVID-19 in the UK from an agreed set of expert academic models. The methods proposed use an equally weighted random effects model, with the REML approach to estimate $\tau^2$, and incorporating either the Wald-type or KNHA approaches for estimating the CIs, to combine estimates from a series of candidate models. 

A meta-analysis using a random effects model as opposed to a fixed effects model is chosen to account for the varying modelling approaches and/or underlying assumptions between candidate models. Moreover, an equally weighted method is adopted in preference to an \textit{inverse-variance} method, as we are combining individual model predictions where additional uncertainty does not necessarily imply imprecision, but is just a reflection of the data being modelled. 

The choice of using the well-established REML to estimate $\tau^2$ is recommended as it has been shown to be robust against deviations from normality - many epidemiological models can, at times, produce skewed output distributions for the parameters of interest.  Both the Wald and KNHA approaches for calculating the CIs perform well, particularly in the case of KNHA, when only a limited number of models are available for comparison\cite{Kontopantelis2012, Kontopantelis2012a, Langan2019}. 

Finally, in order to further protect against skew in the input distributions, an appropriate assessment of the skewed parameters is obtained via optimisation and passed to the \texttt{rma} call from the \texttt{metafor} package\cite{Viechtbauer2010}, together with the estimates from the fitted distributions of each candidate model. The REML method is applied to estimate the heterogeneity variance parameter, and using either the standard Wald-type or KNHA approach for the calculation of the CIs thus enables for an appropriate combined estimates to be formulated.

\section{Acknowledgements}

This work was conducted within a wider effort to improve the policy response to COVID-19. The authors would like to thank the SPI-M modelling groups for the data used in this paper (see Appendix for the complete list of modelling groups). We would also like to thank Tom Finnie at PHE and the SPI-M secretariat for their helpful discussions.

\vspace{1.5cm}

(c) Crown copyright (2021), Dstl. This material is licensed under the terms of the Open Government License except where otherwise stated. To view this license, visit \href{http://www.nationalarchives.gov.uk/doc/open-government-licence/version/3}{\color{blue}http://www.nationalarchives.gov.uk/doc/open-government-licence/version/3} or write to the Information Policy Team, The National Archives, Kew, London TW9 4DU, or email: \href{mailto:psi@nationalarchives.gsi.gov.uk}{\color{blue}psi@nationalarchives.gsi.gov.uk}

\clearpage

\bibliography{Bib}

\begin{thebibliography}{52}
\providecommand{\natexlab}[1]{#1}
\providecommand{\url}[1]{\texttt{#1}}
\expandafter\ifx\csname urlstyle\endcsname\relax
  \providecommand{\doi}[1]{doi: #1}\else
  \providecommand{\doi}{doi: \begingroup \urlstyle{rm}\Url}\fi

\bibitem[Wallinga and Teunis(2004)]{Wallinga2004}
Jacco Wallinga and Peter Teunis.
\newblock {Different Epidemic Curves for Severe Acute Respiratory Syndrome
  Reveal Similar Impacts of Control Measures}.
\newblock \emph{American Journal of Epidemiology}, 160\penalty0 (6):\penalty0
  509--516, 09 2004.
\newblock ISSN 0002-9262.
\newblock \doi{10.1093/aje/kwh255}.
\newblock URL \url{https://doi.org/10.1093/aje/kwh255}.

\bibitem[{The Royal Society}({2020 [Last Accessed 06.10.2020]})]{RoyalSoc2020}
{The Royal Society}.
\newblock {Reproduction number (R) and growth rate (r) of the COVID-19 epidemic
  in the UK: methods of estimation, data sources, causes of heterogeneity, and
  use as a guide in policy formulation}, August {2020 [Last Accessed
  06.10.2020]}.
\newblock URL
  \url{https://royalsociety.org/-/media/policy/projects/set-c/set-covid-19-R-estimates.pdf?la=en-GB&hash=FDFFC11968E5D247D8FF641930680BD6}.
\newblock [online].

\bibitem[Wallinga and Lipsitch(2007)]{Wallinga2007}
J~Wallinga and M~Lipsitch.
\newblock {How generation intervals shape the relationship between growth rates
  and reproductive numbers}.
\newblock \emph{Proceedings of the Royal Society B: Biological Sciences},
  274\penalty0 (1609):\penalty0 599--604, 2007.
\newblock \doi{10.1098/rspb.2006.3754}.
\newblock URL
  \url{https://royalsocietypublishing.org/doi/abs/10.1098/rspb.2006.3754}.

\bibitem[Flaxman et~al.(2020)Flaxman, Mishra, Gandy, Unwin, Mellan, Coupland,
  Whittaker, Zhu, Berah, Eaton, Monod, Perez-Guzman, Schmit, Cilloni, Ainslie,
  Baguelin, Boonyasiri, Boyd, Cattarino, Cooper, Cucunubá, Cuomo-Dannenburg,
  Dighe, Djaafara, Dorigatti, van Elsland, FitzJohn, Gaythorpe, Geidelberg,
  Grassly, Green, Hallett, Hamlet, Hinsley, Jeffrey, Knock, Laydon,
  Nedjati-Gilani, Nouvellet, Parag, Siveroni, Thompson, Verity, Volz, Walters,
  Wang, Wang, Watson, Winskill, Xi, Walker, Ghani, Donnelly, Riley, Vollmer,
  Ferguson, Okell, Bhatt, and Team]{Flaxman2020}
Seth Flaxman, Swapnil Mishra, Axel Gandy, H.~Juliette~T. Unwin, Thomas~A.
  Mellan, Helen Coupland, Charles Whittaker, Harrison Zhu, Tresnia Berah,
  Jeffrey~W. Eaton, Mélodie Monod, Pablo~N. Perez-Guzman, Nora Schmit, Lucia
  Cilloni, Kylie E.~C. Ainslie, Marc Baguelin, Adhiratha Boonyasiri, Olivia
  Boyd, Lorenzo Cattarino, Laura~V. Cooper, Zulma Cucunubá, Gina
  Cuomo-Dannenburg, Amy Dighe, Bimandra Djaafara, Ilaria Dorigatti, Sabine~L.
  van Elsland, Richard~G. FitzJohn, Katy A.~M. Gaythorpe, Lily Geidelberg,
  Nicholas~C. Grassly, William~D. Green, Timothy Hallett, Arran Hamlet, Wes
  Hinsley, Ben Jeffrey, Edward Knock, Daniel~J. Laydon, Gemma Nedjati-Gilani,
  Pierre Nouvellet, Kris~V. Parag, Igor Siveroni, Hayley~A. Thompson, Robert
  Verity, Erik Volz, Caroline~E. Walters, Haowei Wang, Yuanrong Wang, Oliver~J.
  Watson, Peter Winskill, Xiaoyue Xi, Patrick~GT Walker, Azra~C. Ghani,
  Christl~A. Donnelly, Steven~M. Riley, Michaela A.~C. Vollmer, Neil~M.
  Ferguson, Lucy~C. Okell, Samir Bhatt, and Imperial College COVID-19~Response
  Team.
\newblock {Estimating the effects of non-pharmaceutical interventions on
  COVID-19 in Europe}.
\newblock \emph{Nature}, pages~--, 2020.
\newblock ISSN 1476-4687.
\newblock URL \url{https://doi.org/10.1038/s41586-020-2405-7}.

\bibitem[Ridenhour et~al.(2014)Ridenhour, Kowalik, and Shay]{Ridenhour2014}
Benjamin Ridenhour, Jessica~M. Kowalik, and David~K. Shay.
\newblock {Unraveling R0: Considerations for Public Health Applications}.
\newblock \emph{American Journal of Public Health}, 104\penalty0 (2):\penalty0
  e32--e41, 2014.
\newblock \doi{10.2105/AJPH.2013.301704}.
\newblock URL \url{https://doi.org/10.2105/AJPH.2013.301704}.
\newblock PMID: 24328646.

\bibitem[Kucharski et~al.(2020)Kucharski, Klepac, Conlan, Kissler, Tang, Fry,
  Gog, Edmunds, Emery, Medley, Munday, Russell, Leclerc, Diamond, Procter,
  Gimma, Sun, Gibbs, Rosello, van Zandvoort, Hué, Meakin, Deol, Knight,
  Jombart, Foss, Bosse, Atkins, Quilty, Lowe, Prem, Flasche, Pearson, Houben,
  Nightingale, Endo, Tully, Liu, Villabona-Arenas, O'Reilly, Funk, Eggo, Jit,
  Rees, Hellewell, Clifford, Jarvis, Abbott, Auzenbergs, Davies, and
  Simons]{Kucharski2020}
Adam~J Kucharski, Petra Klepac, Andrew J~K Conlan, Stephen~M Kissler, Maria~L
  Tang, Hannah Fry, Julia~R Gog, W~John Edmunds, Jon~C Emery, Graham Medley,
  James~D Munday, Timothy~W Russell, Quentin~J Leclerc, Charlie Diamond,
  Simon~R Procter, Amy Gimma, Fiona~Yueqian Sun, Hamish~P Gibbs, Alicia
  Rosello, Kevin van Zandvoort, Stéphane Hué, Sophie~R Meakin, Arminder~K
  Deol, Gwen Knight, Thibaut Jombart, Anna~M Foss, Nikos~I Bosse, Katherine~E
  Atkins, Billy~J Quilty, Rachel Lowe, Kiesha Prem, Stefan Flasche, Carl A~B
  Pearson, Rein M G~J Houben, Emily~S Nightingale, Akira Endo, Damien~C Tully,
  Yang Liu, Julian Villabona-Arenas, Kathleen O'Reilly, Sebastian Funk,
  Rosalind~M Eggo, Mark Jit, Eleanor~M Rees, Joel Hellewell, Samuel Clifford,
  Christopher~I Jarvis, Sam Abbott, Megan Auzenbergs, Nicholas~G Davies, and
  David Simons.
\newblock {Effectiveness of isolation, testing, contact tracing, and physical
  distancing on reducing transmission of SARS-CoV-2 in different settings: a
  mathematical modelling study}.
\newblock \emph{The Lancet Infectious Diseases}, pages~--, 2020.
\newblock ISSN 1473-3099.
\newblock \doi{10.1016/S1473-3099(20)30457-6}.
\newblock URL \url{https://doi.org/10.1016/S1473-3099(20)30457-6}.

\bibitem[Farrington et~al.(2001)Farrington, Kanaan, and Gay]{Farrington2001}
C.~P. Farrington, M.~N. Kanaan, and N.~J. Gay.
\newblock {Estimation of the basic reproduction number for infectious diseases
  from age-stratified serological survey data}.
\newblock \emph{Journal of the Royal Statistical Society: Series C (Applied
  Statistics)}, 50\penalty0 (3):\penalty0 251--292, 2001.
\newblock \doi{10.1111/1467-9876.00233}.
\newblock URL
  \url{https://rss.onlinelibrary.wiley.com/doi/abs/10.1111/1467-9876.00233}.

\bibitem[DHSC(2020)]{DHSC2020}
DHSC.
\newblock {The R value and growth rate}.
\newblock Website, May 2020.
\newblock URL \url{https://www.gov.uk/guidance/the-r-value-and-growth-rate}.

\bibitem[Lindström et~al.(2015)Lindström, Tildesley, and Webb]{Lindstrom2015}
Tom Lindström, Michael Tildesley, and Colleen Webb.
\newblock {A Bayesian Ensemble Approach for Epidemiological Projections}.
\newblock \emph{PLOS Computational Biology}, 11\penalty0 (4):\penalty0 1--30,
  04 2015.
\newblock \doi{10.1371/journal.pcbi.1004187}.
\newblock URL \url{https://doi.org/10.1371/journal.pcbi.1004187}.

\bibitem[Ray and Reich(2018)]{Ray2018}
Evan~L. Ray and Nicholas~G. Reich.
\newblock {Prediction of infectious disease epidemics via weighted density
  ensembles}.
\newblock \emph{PLOS Computational Biology}, 14\penalty0 (2):\penalty0 1--23,
  02 2018.
\newblock \doi{10.1371/journal.pcbi.1005910}.
\newblock URL \url{https://doi.org/10.1371/journal.pcbi.1005910}.

\bibitem[Wolpert(1992)]{Wolpert1992}
David~H. Wolpert.
\newblock {Stacked generalization}.
\newblock \emph{Neural Networks}, 5\penalty0 (2):\penalty0 241--259, 1992.
\newblock ISSN 0893-6080.
\newblock \doi{https://doi.org/10.1016/S0893-6080(05)80023-1}.
\newblock URL
  \url{https://www.sciencedirect.com/science/article/pii/S0893608005800231}.

\bibitem[Genest and Zidek(1986)]{Genest1986}
Christian Genest and James~V. Zidek.
\newblock {Combining Probability Distributions: A Critique and an Annotated
  Bibliography}.
\newblock \emph{Statistical Science}, 1\penalty0 (1):\penalty0 114 -- 135,
  1986.
\newblock \doi{10.1214/ss/1177013825}.
\newblock URL \url{https://doi.org/10.1214/ss/1177013825}.

\bibitem[A. et~al.(2006)A., C.E., A., J.R., P.H., D.J., J.E., and
  T.]{OHagan2006}
O'Hagan A., Buck C.E., Daneshkhah A., Eiser J.R., Garthwaite P.H., Jenkinson
  D.J., Oakley J.E., and Rakow T.
\newblock \emph{{Uncertain Judgements: Eliciting Experts' Probabilities}},
  chapter~9, pages 179--192.
\newblock John Wiley \& Sons, Ltd, 2006.
\newblock ISBN 9780470033319.
\newblock \doi{https://doi.org/10.1002/0470033312.ch9}.
\newblock URL
  \url{https://onlinelibrary.wiley.com/doi/abs/10.1002/0470033312.ch9}.

\bibitem[McAndrew et~al.(2021)McAndrew, Wattanachit, Gibson, and
  Reich]{McAndrew2021}
Thomas McAndrew, Nutcha Wattanachit, Graham~C. Gibson, and Nicholas~G. Reich.
\newblock {Aggregating predictions from experts: A review of statistical
  methods, experiments, and applications}.
\newblock \emph{WIREs Computational Statistics}, 13\penalty0 (2):\penalty0
  e1514, 2021.
\newblock \doi{https://doi.org/10.1002/wics.1514}.
\newblock URL \url{https://onlinelibrary.wiley.com/doi/abs/10.1002/wics.1514}.

\bibitem[Ratcliff(1979)]{Ratcliff1979}
R.~Ratcliff.
\newblock {Group reaction time distributions and an analysis of distribution
  statistics.}
\newblock \emph{Psychological bulletin}, 86:\penalty0 446--61, May 1979.

\bibitem[Thomas and Ross(1980)]{Thomas1980}
Ewart A.~C. Thomas and B.~Ross.
\newblock {On appropriate procedures for combining probability distributions
  within the same family}.
\newblock \emph{Journal of Mathematical Psychology}, 21:\penalty0 136--152,
  1980.

\bibitem[Keeney and Raiffa(1976)]{Keeney1976}
Ralph~L. Keeney and Howard Raiffa.
\newblock \emph{{Decisions with multiple objectives: Preferences and value
  tradeoffs.}}
\newblock Cambridge University Press, New York, NY, US, 1976.

\bibitem[Cooke(1991)]{Cooke1991}
Roger Cooke.
\newblock \emph{{Experts in uncertainty: Opinion and subjective probability in
  science.}}
\newblock Environmental ethics and science policy series. Oxford University
  Press, New York, NY, US, 1991.

\bibitem[{Berrocal} et~al.(2007){Berrocal}, {Raftery}, and
  {Gneiting}]{Berrocal2007}
Veronica~J. {Berrocal}, Adrian~E. {Raftery}, and Tilmann {Gneiting}.
\newblock {Combining Spatial Statistical and Ensemble Information in
  Probabilistic Weather Forecasts}.
\newblock \emph{Monthly Weather Review}, 135\penalty0 (4):\penalty0 1386,
  January 2007.
\newblock \doi{10.1175/MWR3341.1}.

\bibitem[{Glahn} et~al.(2009){Glahn}, {Peroutka}, {Wiedenfeld}, {Wagner},
  {Zylstra}, {Schuknecht}, and {Jackson}]{Glahn2009}
Bob {Glahn}, Matthew {Peroutka}, Jerry {Wiedenfeld}, John {Wagner}, Greg
  {Zylstra}, Bryan {Schuknecht}, and Bryan {Jackson}.
\newblock {MOS Uncertainty Estimates in an Ensemble Framework}.
\newblock \emph{Monthly Weather Review}, 137\penalty0 (1):\penalty0 246,
  January 2009.
\newblock \doi{10.1175/2008MWR2569.1}.

\bibitem[{Kleiber} et~al.(2011){Kleiber}, {Raftery}, {Baars}, {Gneiting},
  {Mass}, and {Grimit}]{Kleiber2011}
William {Kleiber}, Adrian~E. {Raftery}, Jeffrey {Baars}, Tilmann {Gneiting},
  Clifford~F. {Mass}, and Eric {Grimit}.
\newblock {Locally Calibrated Probabilistic Temperature Forecasting Using
  Geostatistical Model Averaging and Local Bayesian Model Averaging}.
\newblock \emph{Monthly Weather Review}, 139\penalty0 (8):\penalty0 2630--2649,
  August 2011.
\newblock \doi{10.1175/2010MWR3511.1}.

\bibitem[Ranjan and Gneiting(2010)]{Ranjan2010}
Roopesh Ranjan and Tilmann Gneiting.
\newblock {Combining probability forecasts}.
\newblock \emph{Journal of the Royal Statistical Society: Series B (Statistical
  Methodology)}, 72\penalty0 (1):\penalty0 71--91, 2010.
\newblock \doi{https://doi.org/10.1111/j.1467-9868.2009.00726.x}.
\newblock URL
  \url{https://rss.onlinelibrary.wiley.com/doi/abs/10.1111/j.1467-9868.2009.00726.x}.

\bibitem[Gneiting and Ranjan(2013)]{Gneiting2013}
Tilmann Gneiting and Roopesh Ranjan.
\newblock {Combining predictive distributions}.
\newblock \emph{Electronic Journal of Statistics}, 7\penalty0 (none):\penalty0
  1747--1782, 2013.
\newblock \doi{10.1214/13-EJS823}.
\newblock URL \url{https://doi.org/10.1214/13-EJS823}.

\bibitem[Silk et~al.(2020)Silk, Bowman, Dalrymple, and Woods]{Silk2020}
D.~S. Silk, V.~E. Bowman, U.~Dalrymple, and D.~C. Woods.
\newblock {Uncertainty quantification for epidemiological forecasts of COVID-19
  through combinations of model predictions}, 2020.

\bibitem[Funk et~al.(2020)Funk, Abbott, Atkins, Baguelin, Baillie, Birrell,
  Blake, Bosse, Burton, Carruthers, Davies, De~Angelis, Dyson, Edmunds, Eggo,
  Ferguson, Gaythorpe, Gorsich, Guyver-Fletcher, Hellewell, Hill, Holmes,
  House, Jewell, Jit, Jombart, Joshi, Keeling, Kendall, Knock, Kucharski,
  Lythgoe, Meakin, Munday, Openshaw, Overton, Pagani, Pearson, Perez-Guzman,
  Pellis, Scarabel, Semple, Sherratt, Tang, Tildesley, Van~Leeuwen, Whittles,
  Group, Team, and Investigators]{Funk2020}
S~Funk, S~Abbott, BD~Atkins, M~Baguelin, JK~Baillie, P~Birrell, J~Blake,
  NI~Bosse, J~Burton, J~Carruthers, NG~Davies, D~De~Angelis, L~Dyson,
  WJ~Edmunds, RM~Eggo, NM~Ferguson, K~Gaythorpe, E~Gorsich, G~Guyver-Fletcher,
  J~Hellewell, EM~Hill, A~Holmes, TA~House, C~Jewell, M~Jit, T~Jombart,
  I~Joshi, MJ~Keeling, E~Kendall, ES~Knock, AJ~Kucharski, KA~Lythgoe,
  SR~Meakin, JD~Munday, PJM Openshaw, CE~Overton, F~Pagani, J~Pearson,
  PN~Perez-Guzman, L~Pellis, F~Scarabel, MG~Semple, K~Sherratt, M~Tang,
  MJ~Tildesley, E~Van~Leeuwen, LK~Whittles, CMMID COVID-19~Working Group,
  Imperial College COVID-19~Response Team, and ISARIC4C Investigators.
\newblock Short-term forecasts to inform the response to the covid-19 epidemic
  in the uk.
\newblock \emph{medRxiv}, 2020.
\newblock \doi{10.1101/2020.11.11.20220962}.
\newblock URL
  \url{https://www.medrxiv.org/content/early/2020/12/04/2020.11.11.20220962}.

\bibitem[Gneiting et~al.(2005)Gneiting, Raftery, Westveld~III, and
  Goldman]{Gneiting2005}
Tilmann Gneiting, Adrian~E Raftery, Anton~H Westveld~III, and Tom Goldman.
\newblock {Calibrated probabilistic forecasting using ensemble model output
  statistics and minimum CRPS estimation}.
\newblock \emph{Monthly Weather Review}, 133\penalty0 (5):\penalty0 1098--1118,
  2005.

\bibitem[Nowotarski and Weron(2015)]{Nowotarski2015}
Jakub Nowotarski and Rafal Weron.
\newblock {Computing electricity spot price prediction intervals using quantile
  regression and forecast averaging}.
\newblock \emph{Computational Statistics}, 30\penalty0 (3):\penalty0 791--803,
  2015.
\newblock ISSN 1613-9658.
\newblock \doi{10.1007/s00180-014-0523-0}.
\newblock URL \url{https://doi.org/10.1007/s00180-014-0523-0}.

\bibitem[Borenstein et~al.(2009)Borenstein, Hedges, Higgins, and
  Rothstein]{Borenstein2009}
Michael Borenstein, Larry~V. Hedges, Julian~P.T. Higgins, and Hannah~R.
  Rothstein.
\newblock \emph{{Introduction to meta-analysis}}.
\newblock Wiley Online Library, Chichester, London, UK, 2009.
\newblock \doi{10.1002/9780470743386}.
\newblock URL
  \url{https://onlinelibrary.wiley.com/doi/book/10.1002/9780470743386}.

\bibitem[Langan et~al.(2019)Langan, Higgins, Jackson, Bowden, Veroniki,
  Kontopantelis, Viechtbauer, and Simmonds]{Langan2019}
Dean Langan, Julian~P.T. Higgins, Dan Jackson, Jack Bowden, Areti~Angeliki
  Veroniki, Evangelos Kontopantelis, Wolfgang Viechtbauer, and Mark Simmonds.
\newblock {A comparison of heterogeneity variance estimators in simulated
  random-effects meta-analyses}.
\newblock \emph{Research Synthesis Methods}, 10\penalty0 (1):\penalty0 83--98,
  2019.
\newblock \doi{10.1002/jrsm.1316}.
\newblock URL \url{https://onlinelibrary.wiley.com/doi/abs/10.1002/jrsm.1316}.

\bibitem[Borenstein et~al.(2010)Borenstein, Hedges, Higgins, and
  Rothstein]{Borenstein2010}
Michael Borenstein, Larry~V. Hedges, Julian~P.T. Higgins, and Hannah~R.
  Rothstein.
\newblock {A basic introduction to fixed-effect and random-effects models for
  meta-analysis}.
\newblock \emph{Research Synthesis Methods}, 1\penalty0 (2):\penalty0 97--111,
  2010.
\newblock \doi{10.1002/jrsm.12}.
\newblock URL \url{https://onlinelibrary.wiley.com/doi/abs/10.1002/jrsm.12}.

\bibitem[Veroniki et~al.(2016)Veroniki, Jackson, Viechtbauer, Bender, Bowden,
  Knapp, Kuss, Higgins, Langan, and Salanti]{Veroniki2016}
Areti~Angeliki Veroniki, Dan Jackson, Wolfgang Viechtbauer, Ralf Bender, Jack
  Bowden, Guido Knapp, Oliver Kuss, Julian~PT Higgins, Dean Langan, and Georgia
  Salanti.
\newblock {Methods to estimate the between-study variance and its uncertainty
  in meta-analysis}.
\newblock \emph{Research Synthesis Methods}, 7\penalty0 (1):\penalty0 55--79,
  2016.
\newblock \doi{10.1002/jrsm.1164}.
\newblock URL \url{https://onlinelibrary.wiley.com/doi/abs/10.1002/jrsm.1164}.

\bibitem[Hartung and Knapp(2001)]{Hartung2001}
J~Hartung and G~Knapp.
\newblock {A refined method for the meta-analysis of controlled clinical trials
  with binary outcome}.
\newblock \emph{Statistics in medicine}, 20\penalty0 (24):\penalty0
  3875—3889, December 2001.
\newblock ISSN 0277-6715.
\newblock \doi{10.1002/sim.1009}.
\newblock URL \url{https://doi.org/10.1002/sim.1009}.

\bibitem[Sidik and Jonkman(2002)]{Sidik2002}
Kurex Sidik and Jeffrey~N. Jonkman.
\newblock {A simple confidence interval for meta-analysis}.
\newblock \emph{Statistics in Medicine}, 21\penalty0 (21):\penalty0 3153--3159,
  2002.
\newblock \doi{10.1002/sim.1262}.
\newblock URL \url{https://onlinelibrary.wiley.com/doi/abs/10.1002/sim.1262}.

\bibitem[Kontopantelis and Reeves(2012{\natexlab{a}})]{Kontopantelis2012}
Evangelos Kontopantelis and David Reeves.
\newblock {Performance of statistical methods for meta-analysis when true study
  effects are non-normally distributed: A simulation study}.
\newblock \emph{Statistical Methods in Medical Research}, 21\penalty0
  (4):\penalty0 409--426, 2012{\natexlab{a}}.
\newblock \doi{10.1177/0962280210392008}.
\newblock URL \url{https://doi.org/10.1177/0962280210392008}.
\newblock PMID: 21148194.

\bibitem[Kontopantelis and Reeves(2012{\natexlab{b}})]{Kontopantelis2012a}
Evangelos Kontopantelis and David Reeves.
\newblock {Performance of statistical methods for meta-analysis when true study
  effects are non-normally distributed: A comparison between
  {DerSimonian-Laird} and restricted maximum likelihood}.
\newblock \emph{Statistical Methods in Medical Research}, 21\penalty0
  (6):\penalty0 657--659, 2012{\natexlab{b}}.
\newblock \doi{10.1177/0962280211413451}.
\newblock URL \url{https://doi.org/10.1177/0962280211413451}.
\newblock PMID: 23171971.

\bibitem[Bowley(1920)]{Bowley1920}
A~L Bowley.
\newblock \emph{{Elements of Statistics}}.
\newblock Scribner, 4th edition, 1920.

\bibitem[Bulmer(1979)]{Bulmer1979}
Michael~George Bulmer.
\newblock \emph{Principles of statistics}.
\newblock Dover publ., 1979.

\bibitem[Kennedy and Eberhart(1995)]{Kennedy1995}
James Kennedy and Russell Eberhart.
\newblock {Particle swarm optimization}.
\newblock In \emph{{Proceedings of ICNN'95-International Conference on Neural
  Networks}}, volume~4, pages 1942--1948. IEEE, 1995.

\bibitem[Bendtsen(2012)]{Bendtsen2012}
Claus Bendtsen.
\newblock \emph{{pso: Particle Swarm Optimization}}, 2012.
\newblock URL \url{https://CRAN.R-project.org/package=pso}.
\newblock R package version 1.0.3.

\bibitem[{R Core Team}(2019)]{RCoreTeam2019}
{R Core Team}.
\newblock \emph{{R: A Language and Environment for Statistical Computing}}.
\newblock R Foundation for Statistical Computing, Vienna, Austria, 2019.
\newblock URL \url{https://www.R-project.org/}.

\bibitem[Yang(2014)]{YANG201499}
Xin-She Yang.
\newblock {Chapter 7 - Particle Swarm Optimization}.
\newblock In Xin-She Yang, editor, \emph{{Nature-Inspired Optimization
  Algorithms}}, pages 99--110. Elsevier, Oxford, 2014.
\newblock ISBN 978-0-12-416743-8.
\newblock \doi{https://doi.org/10.1016/B978-0-12-416743-8.00007-5}.
\newblock URL
  \url{https://www.sciencedirect.com/science/article/pii/B9780124167438000075}.

\bibitem[Bonett(2008)]{Bonett2008}
Douglas~G. Bonett.
\newblock {Meta-analytic interval estimation for bivariate correlations}.
\newblock \emph{Psychological Methods}, 13\penalty0 (3):\penalty0 173--181,
  2008.
\newblock ISSN 1939-1463(Electronic),1082-989X(Print).
\newblock \doi{10.1037/a0012868}.

\bibitem[Viechtbauer(2010)]{Viechtbauer2010}
Wolfgang Viechtbauer.
\newblock {Conducting meta-analyses in R with the metafor package}.
\newblock \emph{Journal of Statistical Software}, 36\penalty0 (3):\penalty0
  1--48, 2010.
\newblock URL \url{https://www.jstatsoft.org/v36/i03/}.

\bibitem[Higgins et~al.(2019)Higgins, Li, and Deeks]{Higgins2019}
Julian~PT Higgins, Tianjing Li, and Jonathan~J Deeks.
\newblock \emph{{Choosing effect measures and computing estimates of effect}},
  chapter~6, pages 143--176.
\newblock John Wiley \& Sons, Ltd, 2019.
\newblock ISBN 9781119536604.
\newblock \doi{10.1002/9781119536604.ch6}.
\newblock URL
  \url{https://onlinelibrary.wiley.com/doi/abs/10.1002/9781119536604.ch6}.

\bibitem[M.(2015)]{Bland2015}
Bland M.
\newblock {Estimating Mean and Standard Deviation from the Sample Size, Three
  Quartiles, Minimum, and Maximum}.
\newblock \emph{International Journal of Statistics in Medical Research},
  4:\penalty0 57--64, 2015.
\newblock \doi{10.6000/1929-6029.2015.04.01.6}.
\newblock URL \url{https://doi.org/10.6000/1929-6029.2015.04.01.6}.

\bibitem[Wan et~al.(2014)Wan, Wang, Liu, and Tong]{Wan2014}
Xiang Wan, Wenqian Wang, Jiming Liu, and Tiejun Tong.
\newblock {Estimating the sample mean and standard deviation from the sample
  size, median, range and/or interquartile range}.
\newblock \emph{BMC Medical Research Methodology}, 14\penalty0 (1):\penalty0
  135, 2014.
\newblock ISSN 1471-2288.
\newblock \doi{10.1186/1471-2288-14-135}.
\newblock URL \url{https://doi.org/10.1186/1471-2288-14-135}.

\bibitem[Bonett(2009)]{Bonett2009}
Douglas~G. Bonett.
\newblock {Meta-analytic interval estimation for standardized and
  unstandardized mean differences.}
\newblock \emph{Psychological methods}, 14:\penalty0 225--38, Sep 2009.

\bibitem[Shuster(2010)]{Shuster2010}
Jonathan~J. Shuster.
\newblock {Empirical vs natural weighting in random effects meta-analysis.}
\newblock \emph{Statistics in medicine}, 29:\penalty0 1259--65, May 2010.

\bibitem[ONS(2020)]{ONS2020}
ONS.
\newblock {COVID-19 Infection Survey (Pilot): methods and further information}.
\newblock Website, July 2020.
\newblock URL
  \url{https://www.ons.gov.uk/peoplepopulationandcommunity/healthandsocialcare/conditionsanddiseases/methodologies/covid19infectionsurveypilotmethodsandfurtherinformation}.

\bibitem[Jarvis et~al.(2020)Jarvis, Van~Zandvoort, Gimma, Prem, Auzenbergs,
  O'Reilly, Medley, Emery, Houben, Davies, Nightingale, Flasche, Jombart,
  Hellewell, Abbott, Munday, Bosse, Funk, Sun, Endo, Rosello, Procter,
  Kucharski, Russell, Knight, Gibbs, Leclerc, Quilty, Diamond, Liu, Jit,
  Clifford, Pearson, Eggo, Deol, Klepac, Rubin, Edmunds, and working
  group]{Jarvis2020}
Christopher~I. Jarvis, Kevin Van~Zandvoort, Amy Gimma, Kiesha Prem, Megan
  Auzenbergs, Kathleen O'Reilly, Graham Medley, Jon~C. Emery, Rein M. G.~J.
  Houben, Nicholas Davies, Emily~S. Nightingale, Stefan Flasche, Thibaut
  Jombart, Joel Hellewell, Sam Abbott, James~D. Munday, Nikos~I. Bosse,
  Sebastian Funk, Fiona Sun, Akira Endo, Alicia Rosello, Simon~R. Procter,
  Adam~J. Kucharski, Timothy~W. Russell, Gwen Knight, Hamish Gibbs, Quentin
  Leclerc, Billy~J. Quilty, Charlie Diamond, Yang Liu, Mark Jit, Samuel
  Clifford, Carl A.~B. Pearson, Rosalind~M. Eggo, Arminder~K. Deol, Petra
  Klepac, G.~James Rubin, W.~John Edmunds, and C.~M. M. I. D. C. O. V. I. D.-19
  working group.
\newblock {Quantifying the impact of physical distance measures on the
  transmission of COVID-19 in the UK}.
\newblock \emph{BMC Medicine}, 18\penalty0 (1):\penalty0 124, 2020.
\newblock ISSN 1741-7015.
\newblock \doi{10.1186/s12916-020-01597-8}.
\newblock URL \url{https://doi.org/10.1186/s12916-020-01597-8}.

\bibitem[London(2021)]{REACT2021}
Imperial~College London.
\newblock {Real-time Assessment of Community Transmission (REACT) Study}.
\newblock Website, July 2021.
\newblock URL
  \url{https://www.imperial.ac.uk/medicine/research-and-impact/groups/react-study/}.

\bibitem[DHSC(2021)]{DHSC2021}
DHSC.
\newblock {Reproduction number (R) and growth rate: methodology}.
\newblock Website, April 2021.
\newblock URL
  \url{https://www.gov.uk/government/publications/reproduction-number-r-and-growth-rate-methodology/reproduction-number-r-and-growth-rate-methodology}.

\end{thebibliography}

\clearpage

\section{Appendix}\label{appendix}

\subsection{List of SPI-M Modellers/Modelling Groups}

\begin{itemize}
\item Dr Paul Birrell (National Infection Service, Public Health England, London, UK)
\item Dr Jonathan Carruthers (National Infection Service, Public Health England, London, UK)
\item Dr Andr\'e Charlett (Centre for Infectious Disease Surveillance and Control, Public Health England, UK)
\item Prof Daniela DeAngelis (Medical Research Council Biostatistics Unit, School of Clinical Medicine, University of Cambridge, UK)
\item Joshua Blake (Medical Research Council Biostatistics Unit, School of Clinical Medicine, University of Cambridge, UK)
\item Prof Matt Keeling (Department of Biological Sciences and Mathematics Institute, University of Warwick, UK)
\item Dr Louise Dyson (School of Life Sciences and Mathematics Institute, University of Warwick, UK)
\item Dr Sebastian Funk (Centre for the Mathematical Modelling of Infectious Diseases, London School of Hygiene and Tropical Medicine, UK)
\item Dr Sam Abbott (Centre for the Mathematical Modelling of Infectious Diseases, London School of Hygiene and Tropical Medicine, UK)
\item Nikos Bosse (Centre for the Mathematical Modelling of Infectious Diseases, London School of Hygiene and Tropical Medicine, UK)
\item Joel Hellewell (Centre for the Mathematical Modelling of Infectious Diseases, London School of Hygiene and Tropical Medicine, UK)
\item Sophie Meakin (Centre for the Mathematical Modelling of Infectious Diseases, London School of Hygiene and Tropical Medicine, UK)
\item James Munday (Centre for the Mathematical Modelling of Infectious Diseases, London School of Hygiene and Tropical Medicine, UK)
\item Katharine Sherratt (Centre for the Mathematical Modelling of Infectious Diseases, London School of Hygiene and Tropical Medicine, UK)
\item Dr Robin Thompson (Mathematical Institute, University of Oxford, UK)
\item Prof John Edmunds (Centre for the Mathematical Modelling of Infectious Diseases, London School of Hygiene and Tropical Medicine, UK)
\item Dr Nicholas Davies (Centre for the Mathematical Modelling of Infectious Diseases, London School of Hygiene and Tropical Medicine, UK)
\item Dr Christopher Jarvis (Centre for the Mathematical Modelling of Infectious Diseases, London School of Hygiene and Tropical Medicine, UK)
\item Amy Gimma (Centre for the Mathematical Modelling of Infectious Diseases, London School of Hygiene and Tropical Medicine, UK)
\item Kevin Van Zandvoort (Centre for the Mathematical Modelling of Infectious Diseases, London School of Hygiene and Tropical Medicine, UK)
\item Prof Neil Ferguson (Medical Research Council Centre for Outbreak Analysis and Modelling, Imperial College London, UK)
\item Dr Marc Baguelin (Medical Research Council Centre for Outbreak Analysis and Modelling, Imperial College London, UK)
\item Dr Lorenzo Pellis (Department of Mathematics, University of Manchester, UK)
\item Dr Thomas House (Department of Mathematics, University of Manchester, UK)
\item Dr Christopher Overton (Department of Mathematics, University of Manchester, UK)
\item Joshua Burton (Department of Mathematics, University of Manchester, UK)
\item Filippo Pagani (Department of Mathematics, University of Manchester, UK)
\item Prof Katrina Lythgoe (Big Data Institute, University of Oxford, UK)
\item Dr Francesca Scarabel (LIAM, Department of Mathematics and Statistics, York University, Canada)
\item Dr Jonathon Read (Centre for Health Informatics, Computing, and Statistics, Lancaster University, UK)
\item Dr Chris Jewell (Lancaster Medical School, Lancaster University, UK)
\item Dr Leon Danon (College of Engineering and Mathematical Sciences, University of Exeter, UK)
\item Dr Robert Challen (College of Engineering and Mathematical Sciences, University of Exeter, UK)
\item Dr Ellen Brooks-Pollock (Population Health Sciences, University of Bristol, Bristol, UK)
\item Dr Nabeil Salama (Marine Scotland Science, Aberdeen, UK)
\end{itemize}

\clearpage

\subsection{Model Descriptions}

Detailed information on the models are available from the Royal Society pre-print\cite{RoyalSoc2020}. However, the following provides a brief summary of the models obtained from the Department of Health and Social Care\cite{DHSC2021}: 

\begin{itemize}
\item The University of Cambridge MRC Biostatistics Unit and Public Health England (PHE) use a deterministic age-structured compartmental model, incorporating data from the number of daily deaths and serology data (primary inputs), combined with school attendance and mobility data.
\item The University of Warwick use a deterministic age-structured compartmental model, with model parameters fitted to epidemiological data including hospital admissions and bed occupancy, intensive care unit admissions, number of daily deaths, serological data and, for some model configurations, the proportion of Pillar 2 tests that are positive.
\item The London School of Hygiene and Tropical Medicine (LSHTM) jointly estimates the trajectory of infections and reproduction number using a renewal equation model and observed delays, with the model fitted to different data streams (in particular: cases and hospitalisations) separately.
\item The MRC Centre for Global Infectious Disease Analysis at Imperial College London uses a stochastic age-structured compartmental model, which includes transmission in care homes. Model parameters are fitted to epidemiological data, including hospital admissions and bed occupancy, intensive care unit admissions, number of daily deaths, Pillar 2 testing, together with REACT community survey and blood donor serological data.
\item The University of Manchester uses a deterministic compartmental model, incorporating data from hospital admissions, hospital and intensive care unit bed occupancy, and hospital deaths.
\item The Scottish Government uses a hierarchical Bayesian mechanistic model developed by Imperial College London, including the bespoke package Epidemia, to estimate the reproduction number. The model incorporates data including the number of daily deaths and contact patterns.
\item Lancaster University uses two approaches to estimate reproduction numbers; the first is an application of the renewal equation method using the EpiEstim library and using data on cases (England, Scotland) and hospital admissions (Northern Ireland, Wales); and the second approach is a meta-population transmission model of infection within and between local authorities incorporating movement data and fitted to case data.
\item The University of Exeter and University of Bristol use a renewal equation model produced using the EpiEstim library. The model uses data on cases and hospital admissions.

\end{itemize}

\clearpage

\subsection{Combined Estimates of \textit{R(t)}}

\begin{table}[!htb]
	\tiny
\setlength\tabcolsep{1.5pt}
\begin{tabular}{@{}lcccccccccccc@{}}
\toprule
 & Region 1 & Region 2 & Region 3 & Region 4 & Region 5 & Region 6 & Region 7 & Region 8 & Region 9 & Region 10 & Region 11 & Region 12\\ \midrule
Model 1 & 0.83 (0.71, 0.96) & 0.49 (0.31, 0.72) & 0.77 (0.71, 0.82) & 0.78 (0.56, 1.04) & 0.70 (0.54, 0.86) & 0.76 (0.57, 1.00) & 0.73 (0.60, 0.87) & 0.86 (0.74, 0.99) & 0.77 (0.72, 0.82) & 0.74 (0.63, 0.87) & 0.83 (0.63, 1.05) & 0.94 (0.52, 1.56)\\
Model 2 & 0.84 (0.76, 0.92) & 0.67 (0.58, 0.74) & 0.78 (0.70, 0.85) & 0.76 (0.59, 0.92) & 0.69 (0.62, 0.81) & 0.64 (0.52, 0.82) & 0.69 (0.59, 0.75) & 0.85 (0.75, 0.97) & 0.79 (0.68, 0.88) & 0.70 (0.62, 0.83) & 0.87 (0.79, 1.04) & 0.64 (0.51, 0.80)\\
Model 3 & 0.81 (0.72, 0.92) & 0.68 (0.53, 0.85) & 0.83 (0.76, 0.91) & 0.85 (0.67, 1.06) & 0.74 (0.61, 0.89) & 0.57 (0.41, 0.79) & 0.76 (0.65, 0.89) & 0.87 (0.77, 0.98) & 0.84 (0.77, 0.91) & 0.74 (0.64, 0.87) & 0.79 (0.66, 0.93) & 0.46 (0.21, 0.88)\\
Model 4 & 0.76 (0.44, 1.12) & 0.64 (0.37, 0.93) & 0.71 (0.43, 1.01) & 0.68 (0.35, 1.22) & 0.69 (0.38, 1.13) & 0.52 (0.29, 0.79) & 0.80 (0.45, 1.23) & 0.82 (0.43, 1.51) & 0.74 (0.43, 1.07) & 0.75 (0.44, 1.14) & 0.56 (0.31, 0.86) & 0.60 (0.33, 0.95)\\
Model 5 & 0.91 (0.90, 0.92) & 0.75 (0.73, 0.77) & 0.82 (0.81, 0.83) & 0.77 (0.76, 0.78) & 0.74 (0.73, 0.75) & 0.60 (0.58, 0.63) & 0.72 (0.71, 0.74) & 0.78 (0.77, 0.80) & 0.80 (0.80, 0.81) & 0.80 (0.79, 0.80)$^\dagger$ & 0.82 (0.80, 0.84) & 0.69 (0.67, 0.71)\\
Model 6 & 0.85 (0.82, 0.88) & 0.84 (0.81, 0.90) & 0.84 (0.83, 0.86) & 0.86 (0.82, 0.95) & 0.82 (0.80, 0.86) & 0.76 (0.71, 0.85) & 0.82 (0.81, 0.86) & 0.83 (0.81, 0.87) & 0.84 (0.83, 0.86) & 0.83 (0.81, 0.87) & 0.88 (0.81, 0.94) & 0.86 (0.79, 1.01)\\
Model 7 & 0.87 (0.68, 1.07) & 0.98 (0.75, 1.25) & NA & NA & 0.80 (0.61, 1.03) & NA & 0.84 (0.66, 1.04) & 1.00 (0.78, 1.23) & 0.98 (0.86, 1.13) & 0.79 (0.62, 0.99) & 0.91 (0.68, 1.15) & NA\\
Model 8 & NA & NA & NA & NA & NA & 0.69 (0.65, 0.72) & NA & NA & NA & NA & NA & NA\\
Model 9 & 0.96 (0.76, 1.20) & 1.03 (0.63, 1.50) & 0.91 (0.82, 1.01) & 1.06 (0.66, 1.60) & 0.95 (0.70, 1.26) & 0.90 (0.56, 1.37) & 0.98 (0.74, 1.26) & 0.98 (0.77, 1.22) & 0.92 (0.82, 1.02) & 0.94 (0.75, 1.16) & 0.99 (0.67, 1.39) & NA\\
Model 10 & 0.86 (0.84, 0.88) & 0.78 (0.77, 0.80) & 0.76 (0.75, 0.77) & 0.75 (0.73, 0.77) & 0.74 (0.73, 0.76) & 0.64 (0.62, 0.66) & 0.79 (0.77, 0.80) & 0.80 (0.79, 0.82) & 0.82 (0.79, 0.84) & 0.83 (0.82, 0.84) & 0.77 (0.75, 0.79) & 0.78 (0.76, 0.79)\\
Model 11 & 0.97 (0.87, 1.06) & 0.79 (0.59, 0.95) & NA & NA & 0.92 (0.79, 1.06) & NA & 0.89 (0.75, 0.99) & 0.92 (0.82, 1.01) & 0.92 (0.86, 0.97) & 0.93 (0.84, 1.03) & 1.01 (0.84, 1.24) & NA\\
Model 12 & 0.77 (0.68, 0.88) & 0.60 (0.48, 0.75) & 0.77 (0.68, 0.87) & 0.76 (0.65, 0.89) & 0.75 (0.64, 0.88) & 0.63 (0.51, 0.76) & 0.79 (0.70, 0.89) & 0.75 (0.65, 0.86) & 0.79 (0.71, 0.88) & 0.76 (0.66, 0.86) & 0.90 (0.78, 1.02) & 0.95 (0.76, 1.16)\\ \midrule
\textit{Inverse-variance}  &  &  &  &  &  &  &  &  &  &  &  & \\
\textit{Weighted} &  &  &  &  &  &  &  &  &  &  &  & \\
\textit{REML alone} &  &  &  &  &  &  &  &  &  &  &  & \\
$\widehat{\theta}  \;  (95\% CI)$ & 0.87 (0.84, 0.89) & 0.75 (0.70, 0.79) & 0.81 (0.78, 0.83) & 0.77 (0.75, 0.78) & 0.76 (0.73, 0.79) & 0.65 (0.62, 0.69) & 0.78 (0.74, 0.81) & 0.82 (0.79, 0.84) & 0.83 (0.81, 0.85) & 0.81 (0.79, 0.83) & 0.83 (0.79, 0.86) & 0.74 (0.69, 0.80)\\ \midrule
\textit{Equally Weighted} &  &  &  &  &  &  &  &  &  &  &  & \\
\textit{REML alone} &  &  &  &  &  &  &  &  &  &  &  & \\
$\widehat{\theta}  \;  (95\% CI)$ & 0.86 (0.80, 0.91) & 0.75 (0.68, 0.82) & 0.80 (0.76, 0.84) & 0.81 (0.71, 0.90) & 0.78 (0.71, 0.84) & 0.67 (0.60, 0.74) & 0.80 (0.74, 0.86) & 0.86 (0.79, 0.94) & 0.84 (0.80, 0.88) & 0.80 (0.75, 0.85) & 0.85 (0.78, 0.91) & 0.74 (0.62, 0.85)\\ \midrule
\textit{Equally Weighted} &  &  &  &  &  &  &  &  &  &  &  & \\
\textit{REML+KNHA} &  &  &  &  &  &  &  &  &  &  &  & \\
$\widehat{\theta}  \;  (95\% CI)$ & 0.86 (0.81, 0.91) & 0.75 (0.66, 0.84) & 0.80 (0.75, 0.85) & 0.81 (0.72, 0.90) & 0.78 (0.72, 0.84) & 0.67 (0.60, 0.74) & 0.80 (0.74, 0.86) & 0.86 (0.78, 0.94) & 0.84 (0.79, 0.89) & 0.80 (0.74, 0.86) & 0.85 (0.78, 0.92) & 0.74 (0.61, 0.87)\\ \midrule
$\tau^2$ & 0.000915 & 0.002856 & 0.001222 & 0.000004 & 0.001020 & 0.001391 & 0.001677 & 0.000598 & 0.000984 & 0.000427 & 0.001534 & 0.003216\\
$(SE)$ & (0.000982) & (0.002833) & (0.001078) & (0.000177) & (0.001180) & (0.001574) & (0.001643) & (0.000765) & (0.000893) & (0.000555) & (0.001861) & (0.004129)\\ \bottomrule
\end{tabular}
	\caption{$R(t)$ estimates (90\% CIs) for anonymised models 1 to 12 for all anonymised UK nation/regions, together with calculated combined estimates using: an \textit{inverse-variance} weighted approach with Wald-type CIs; an equally weighted approach with Wald-type CIs (\textit{REML alone}); and an equally weighted approach with KNHA CIs (\textit{REML+KNHA}). All numbers displayed to two decimal places except $\tau^2 (SE)$, displayed to six decimal places. Missing values indicate instances where estimates were not available for models for the specific nation/region. $^\dagger$ Estimates found to be moderately to highly skewed.}
\label{appendix:r_data}
\end{table}

\subsection{Combined Estimates of \textit{r(t)}}

\begin{table}[!htb]
	\tiny
\setlength\tabcolsep{1pt}
\begin{tabular}{@{}lcccccccccccc@{}}
\toprule
 & Region 1 & Region 2 & Region 3 & Region 4 & Region 5 & Region 6 & Region 7 & Region 8 & Region 9 & Region 10 & Region 11 & Region 12\\ \midrule
Model 1 & NA & NA & NA & NA & NA & NA & NA & NA & NA & NA & NA & NA\\
Model 2 & NA & NA & NA & NA & NA & NA & NA & NA & NA & NA & NA & NA\\
Model 3 & NA & NA & NA & NA & NA & NA & NA & NA & NA & NA & NA & NA\\
Model 4 & NA & NA & NA & NA & NA & NA & NA & NA & NA & NA & NA & NA\\
Model 5 & -0.01 (-0.01, -0.01) & -0.04 (-0.04, -0.04) & -0.03 (-0.03, -0.02) & -0.04 (-0.04, -0.03) & -0.04 (-0.04, -0.04) & -0.06 (-0.07, -0.06) & -0.04 (-0.05, -0.04) & -0.03 (-0.03, -0.03) & -0.03 (-0.03, -0.03) & -0.03 (-0.03, -0.03) & -0.03 (-0.03, -0.02) & -0.05 (-0.05, -0.05)\\
Model 6 & -0.03 (-0.04, -0.02) & -0.03 (-0.04, -0.02) & -0.03 (-0.03, -0.03) & -0.03 (-0.04, -0.01) & -0.04 (-0.04, -0.03) & -0.05 (-0.06, -0.03) & -0.03 (-0.04, -0.03) & -0.03 (-0.04, -0.03) & -0.03 (-0.03, -0.03) & -0.03 (-0.04, -0.02) & -0.02 (-0.04, -0.01) & -0.03 (-0.04,  0.00)\\
Model 7 & -0.03 (-0.07,  0.02) & -0.01 (-0.06,  0.05) & NA & NA & -0.05 (-0.10,  0.00) & NA & -0.04 (-0.09,  0.01) &  0.00 (-0.05,  0.04) & NA & -0.05 (-0.10, -0.01) & -0.02 (-0.08,  0.03) & NA\\
Model 8 & NA & NA & NA & NA & NA & NA & NA & NA & NA & NA & NA & NA\\
Model 9 & -0.01 (-0.07,  0.06) &  0.01 (-0.11,  0.14) & -0.02 (-0.06,  0.00) &  0.02 (-0.10,  0.17) & -0.01 (-0.09,  0.08) & -0.03 (-0.14,  0.11) & -0.01 (-0.08,  0.07) & -0.01 (-0.07,  0.06) & -0.02 (-0.06,  0.01) & -0.02 (-0.07,  0.04) &  0.00 (-0.10,  0.11) & NA\\
Model 10 & -0.03 (-0.03, -0.02) & -0.03 (-0.04, -0.03) & -0.04 (-0.04, -0.04) & -0.04 (-0.04, -0.04) & -0.04 (-0.04, -0.04) & -0.06 (-0.06, -0.05) & -0.03 (-0.04, -0.03) & -0.03 (-0.04, -0.03) & -0.03 (-0.04, -0.03) & -0.03 (-0.04, -0.03) & -0.04 (-0.04, -0.04) & -0.03 (-0.04, -0.03)\\
Model 11 & -0.01 (-0.04,  0.02) & -0.06 (-0.13, -0.01) & NA & NA & -0.02 (-0.06,  0.02) & NA & -0.03 (-0.07,  0.00) & -0.02 (-0.06,  0.00) & -0.02 (-0.04, -0.01) & -0.02 (-0.05,  0.01) &  0.00 (-0.05,  0.07) & NA\\
Model 12 & NA & NA & NA & NA & NA & NA & NA & NA & NA & NA & NA & NA\\ \midrule
\textit{Inverse-variance} &  &  &  &  &  &  &  &  &  &  &  & \\
\textit{Weighted} &  &  &  &  &  &  &  &  &  &  &  & \\
\textit{REML alone} &  &  &  &  &  &  &  &  &  &  &  & \\
$\widehat{\theta}  \;  (95\% CI)$ & -0.02 (-0.03, -0.01) & -0.04 (-0.04, -0.03) & -0.03 (-0.04, -0.03) & -0.04 (-0.04, -0.03) & -0.04 (-0.04, -0.04) & -0.06 (-0.07, -0.05) & -0.04 (-0.04, -0.03) & -0.03 (-0.03, -0.03) & -0.03 (-0.03, -0.03) & -0.03 (-0.03, -0.03) & -0.03 (-0.04, -0.02) & -0.04 (-0.05, -0.03)\\ \midrule
\textit{Equally Weighted} &  &  &  &  &  &  &  &  &  &  &  & \\
\textit{REML alone} &  &  &  &  &  &  &  &  &  &  &  & \\
$\widehat{\theta}  \;  (95\% CI)$ & -0.02 (-0.04, 0.00) & -0.03 (-0.05, 0.00) & -0.03 (-0.04, -0.02) & -0.02 (-0.06, 0.02) & -0.03 (-0.05, -0.01) & -0.05 (-0.08, -0.01) & -0.03 (-0.05, -0.01) & -0.02 (-0.04, -0.01) & -0.03 (-0.04, -0.02) & -0.03 (-0.04, -0.02) & -0.02 (-0.04, 0.01) & -0.04 (-0.05, -0.02)\\ \midrule
\textit{Equally Weighted} &  &  &  &  &  &  &  &  &  &  &  & \\
\textit{REML+KNHA} &  &  &  &  &  &  &  &  &  &  &  & \\
$\widehat{\theta}  \;  (95\% CI)$ & -0.02 (-0.03, 0.00) & -0.03 (-0.05, 0.00) & -0.03 (-0.04, -0.02) & -0.02 (-0.07, 0.03) & -0.03 (-0.05, -0.02) & -0.05 (-0.09, -0.01) & -0.03 (-0.05, -0.02) & -0.02 (-0.03, -0.01) & -0.03 (-0.03, -0.02) & -0.03 (-0.04, -0.02) & -0.02 (-0.04, 0.00) & -0.04 (-0.06, -0.02)\\ \midrule
$\tau^2$ & 0.000062 & 0.000009 & 0.000034 & 0.000004 & 0.000000 & 0.000019 & 0.000023 & 0.000000 & 0.000000 & 0.000004 & 0.000076 & 0.000084\\
$(SE)$ & (0.000067) & (0.000016) & (0.000036) & (0.000009) & (0.000003) & (0.000037) & (0.000029) & (0.000004) & (0.000003) & (0.000007) & (0.000093) & (0.000120)\\ \bottomrule
\end{tabular}
	\caption{$r(t)$ estimates (90\% CIs) for anonymised models 1 to 12 for all anonymised UK nation/regions, together with calculated combined estimates using: an \textit{inverse-variance} weighted approach with Wald-type CIs; an equally weighted approach with Wald-type CIs (\textit{REML alone}); and an equally weighted approach with KNHA CIs (\textit{REML+KNHA}). All numbers displayed to two decimal places except $\tau^2 (SE)$, displayed to six decimal places. Missing values indicate instances where estimates were not available for models for the specific nation/region. }
\label{appendix:gr_data}
\end{table}

\clearpage

\end{document}